\documentclass[prd,reprint,superscriptaddress,showpacs,showkeys]{revtex4-1}

\usepackage[utf8]{inputenc}
\usepackage[T1]{fontenc}
\usepackage[english]{babel}
\usepackage{amsfonts}
\usepackage{amssymb}
\usepackage{amsmath}
\usepackage{cancel}
\usepackage{dsfont}
\usepackage{bm}
\usepackage{braket}
\usepackage{simplewick}
\usepackage{slashed}
\usepackage[version=4]{mhchem}
\usepackage{verbatim}
\usepackage{enumitem}
\usepackage{booktabs}
\usepackage{array}
\usepackage{dcolumn}
\usepackage{longtable}
\usepackage{rotating}
\usepackage{multirow}
\usepackage{graphicx}
\usepackage{xcolor}
\usepackage[colorlinks]{hyperref}

\newcolumntype{d}[1]{D{.}{.}{#1}}

\definecolor{color_01}{rgb}{0,0,0.75}
\definecolor{color_02}{rgb}{0.75,0,0}
\definecolor{color_03}{rgb}{0.5,0,0.75}
\hypersetup{linkcolor=color_01,citecolor=color_02,urlcolor=color_03}

\begin{document}
\title{Quark condensate seesaw mechanism for neutrino mass}

\author{A.~Babi\v{c}}
\affiliation{Faculty of Nuclear Sciences and Physical Engineering, Czech Technical University in Prague, 115~19~Prague, Czech Republic}
\affiliation{Institute of Experimental and Applied Physics, Czech Technical University in Prague, 110~00~Prague, Czech Republic}
\affiliation{Bogoliubov Laboratory of Theoretical Physics, Joint Institute for Nuclear Research, 141980~Dubna, Russia}
\author{S.~Kovalenko}
\affiliation{Departamento de Ciencias F\'{i}sicas, Universidad Andres Bello, Sazie~2212, Piso~7, Santiago, Chile}
\affiliation{Centro Cient\'{i}fico Tecnol\'{o}gico de Valpara\'{i}so, Casilla~110-V, Valpara\'{i}so, Chile}
\author{M.~I.~Krivoruchenko}
\affiliation{Institute for Theoretical and Experimental Physics, NRC ``Kurchatov Institute,'' B.~Cheremushkinskaya~25, 117218~Moscow, Russia}
\affiliation{National Research Center ``Kurchatov Institute,'' Ploshchad' Akademika Kurchatova~1, 123182~Moscow, Russia}
\author{F.~\v{S}imkovic}
\affiliation{Institute of Experimental and Applied Physics, Czech Technical University in Prague, 110~00~Prague, Czech Republic}
\affiliation{Bogoliubov Laboratory of Theoretical Physics, Joint Institute for Nuclear Research, 141980~Dubna, Russia}
\affiliation{Faculty of Mathematics, Physics and Informatics, Comenius University in Bratislava, 842~48~Bratislava, Slovakia}

\date{\today}

\begin{abstract}
We study a mechanism of generation of Majorana neutrino mass due to spontaneous breaking of chiral symmetry (SBCS) accompanied by the formation of a quark condensate. The effect of the condensate 
is transmitted to the neutrino sector via Lepton-Number Violating (LNV) lepton-quark dimension-$7$ operators known in the literature as an origin of the neutrino-mass-independent mechanism of neutrinoless double-beta ($0 \nu \beta \beta$) decay. The smallness of neutrino masses is due to a large ratio between the LNV scale and the scale of the SBCS. This is a new realization of the seesaw mechanism, which we dub the Quark Condensate SeeSaw (QCSS). 
%
We examine the predictions of the QCSS for $0 \nu \beta \beta$-decay and neutrino 
mass spectrum. We will show that our model predicts the normal neutrino mass ordering and narrow ranges of the neutrino masses. 
\end{abstract}


\maketitle

\section{Introduction}
The smallness of neutrino masses, in comparison with the other Standard Model (SM) 
fermions, remains a mystery 
of particle physics theory. A common wisdom suggests that this smallness is related to some broken symmetry. 
One of the most natural candidates is $\mathrm{U}(1)_L$ symmetry of the Lepton (L) Number, broken at some high-energy scale $\Lambda$.
Then at the electroweak scale there appears the $\Delta L = 2$ Weinberg operator
\begin{equation}
\label{eq:Weinberg-1}
\mathcal{O}_W = \frac{f}{\Lambda} \, \overline{L^C} \, H \, L \, H,
\end{equation}
which, after the electroweak symmetry breaking (EWSB) $\sqrt{2} \langle H^0 \rangle = v = 246 \, \mathrm{GeV}$, leads to the Majorana neutrino mass
\begin{equation}
\label{eq:M-nu-10}
m_{\nu} = -f \, v \, \frac{v}{\Lambda}.
\end{equation}
For a generic case with $f \sim 1$ and for $m_{\nu}$ at a sub-$\mathrm{eV}$ scale one estimates $\Lambda \sim 10^{14\textnormal{--}15} \, \mathrm{GeV}$, putting Lepton Number Violating (LNV) physics far beyond the experimental reach. This happens in the tree-level realizations of the Weinberg operator (\ref{eq:Weinberg-1}) in the celebrated seesaw Type I, II and III, where $\Lambda$ is equal to the masses $M$ of the corresponding seesaw messengers which, being very heavy, have no phenomenological significance. In order to escape this situation and open up the possibility for a non-trivial phenomenology, various
models have been proposed in the literature (for a recent review see, e.g., Ref.~\cite{Cai:2017jrq}) relaxing the above-mentioned limitation on the LNV scale $\Lambda$. Introducing new symmetries (softly-broken), one can forbid the operator  (\ref{eq:Weinberg-1}) at the tree level, while allowing it at certain loop level $l$, so that in (\ref{eq:M-nu-10}) there appears  a loop suppression factor $f \sim (1/16 \pi^2)^l$.
With the appropriate $l$ the LNV scale $\Lambda$ can be reduced down to phenomenologically interesting values in the $\mathrm{TeV}$ ballpark (see, e.g., Refs.~\cite{Ma:2016mwh,Yao:2017vtm,CentellesChulia:2018bkz,CentellesChulia:2019xky,Arbelaez:2019wyz,Arbelaez:2019ofg,Babu20} and references therein).
Another possibility is to resort to symmetries forbidding (\ref{eq:Weinberg-1}) at all, but allowing higher dimension-$(5 + n)$ operators which
%
after EWSB provide an extra suppression factor $(v/\Lambda)^n$.
As in the loop-based models, here, for sufficiently large $n$, the LNV scale $\Lambda$ can be made as low as the current experimental limits. In some models both loop and higher-dimension suppressions can be combined.

In the present paper we consider another class of the SM gauge-invariant effective operators
\begin{equation}
\label{eq:Eff-Oper-dim-7-1} 
\mathcal{O}_7^{u, d} = \frac{g_{\alpha \beta}^{u, d}}{\Lambda^3} \, \overline{L_{\alpha}^C} \, L_{\beta} \, H \left\lbrace (\overline{Q} \, u_R), \, (\overline{d_R} \, Q)\right\rbrace.
%
%
\end{equation}
Here, all the possible $\mathrm{SU}(2)_L$ contractions are assumed. 
The operators (\ref{eq:Eff-Oper-dim-7-1}) were previously studied in the literature as a source of $\Delta L = 2$ interactions able to induce $0 \nu \beta \beta$-decay with no explicit dependence on the Majorana neutrino mass
\cite{Pas:1999fc,Deppisch:2012nb,Arbelaez:2016zlt,Cirigliano:2017djv}.
On the other hand, it was observed in Ref.~\cite{Tho92} that this operator contributes to the Majorana-neutrino mass matrix due to spontaneous breaking of chiral symmetry (SBCS) 
via the light-quark condensate $\langle \overline{q} q \rangle = -\omega^3 \ne 0$. The latter  sets the SBCS scale, so that after the EWSB and SBCS one arrives at the contribution to the Majorana mass matrix of active neutrinos 
\begin{equation}
\label{eq:Dim-7-Mnu}
m_{\alpha \beta}^{\nu} = -\frac{g_{\alpha \beta}}{\sqrt{2}} \, v \, \frac{\langle \overline{q} q \rangle}{\Lambda^3} = \frac{g_{\alpha \beta}}{\sqrt{2}} \, v \left( \frac{\omega}{\Lambda} \right)^3,
\end{equation}
with $g_{\alpha \beta} = g_{\alpha \beta}^u + g_{\alpha \beta}^d$, where $\langle \overline{q} q \rangle \equiv \langle \overline{u} u \rangle \approx \langle \overline{d} d \rangle \approx 2 \, \langle \overline{u_L} \, u_R \rangle \approx 2 \, \langle \overline{d_R} \, d_L \rangle$. This is a kind of seesaw formula relating the smallness of the Majorana masses of neutrino with the large ratio between the scale $\Lambda$ of Lepton-Number Violation (LNV) and the scale of chiral-symmetry breaking $\omega = -\langle \overline{q} q \rangle^{1/3}$. We call the relation (\ref{eq:Dim-7-Mnu}) \emph{Quark Condensate Seesaw (QCSS) formula}. Taking
\begin{equation}
\label{eq:quark-condensate-value}
\langle \overline{q} q \rangle^{1/3} \approx -283 \, \mathrm{MeV}
\end{equation}
from a renormalized lattice QCD within the $\overline{\mathrm{MS}}$ scheme at a fixed scale $\mu = 2 \, \mathrm{GeV}$ \cite{McNeile:2012xh} and $\Lambda \sim$ a few $\mathrm{TeV}$ we get the neutrino mass in the sub-$\mathrm{eV}$ ballpark.

In the next section we study implications of the requirement of the dominance of the operator (\ref{eq:Eff-Oper-dim-7-1}) for UltraViolet (UV) model building and certain phenomenological aspects of the QCSS.
%
In Section~\ref{sec:lim-eps} we extract limits on the couplings of nonstandard neutrino-quark contact interactions appearing in QCSS.
 Then, we analyze contributions of the operators (\ref{eq:Eff-Oper-dim-7-1}) to neutrinoless double-beta ($0 \nu \beta \beta$) decay and derive strong limitations on the QCSS mechanism from this LNV process.

\section{Dominance of QCSS and Light-Quark Masses}
\label{sec:Possible Dominance of QCSS}
Here we discuss the conditions for the dominance of the operator (\ref{eq:Eff-Oper-dim-7-1}) in the Majorana neutrino mass matrix. As usual, this can be guaranteed by imposing on the theory an appropriate symmetry group $\mathcal{G}$ which could be either continuous or discrete. General properties of this kind of symmetries were studied in Ref.~\cite{Tho92}. This symmetry must forbid the Weinberg operator (\ref{eq:Weinberg-1}), but allow the operator $\mathcal{O}_7^q$ in Eq.~(\ref{eq:Eff-Oper-dim-7-1}). Therefore, the lepton bilinear $LL$ must be a $\mathcal{G}$ non-singlet. Requiring that $\mathcal{G}$ remains a good symmetry after the EWSB and still forbids any contribution to the Majorana-neutrino mass term
\begin{equation}
\label{eq:Nu-MassTerm-1}
\mathcal{L}_{\mathrm{M}} = -\frac{1}{2} \sum_{\alpha \beta} \overline{\nu_{\alpha L}^C} \, m_{\alpha \beta}^{\nu} \, \nu_{\beta L} + \mathrm{H.c.}
\end{equation}
while allowing the quark-lepton coupling
\begin{equation}
\label{eq:neutrino-quark coupling}
\mathcal{L}_7 = \frac{1}{\sqrt{2}} \sum_{\alpha \beta} \frac{v}{\Lambda^3} \, \overline{\nu_{\alpha L}^C} \, \nu_{\beta L} \left( g_{\alpha \beta}^u \, \overline{u_L} \, u_R + g_{\alpha \beta}^d \, \overline{d_R} \, d_L \right) + \mathrm{H.c.}
\end{equation}
implies that we claim the SM Higgs $H$ to be a $\mathcal{G}$-singlet. Thus, the condition of $\mathcal{G}$ invariance of the operator (\ref{eq:Eff-Oper-dim-7-1}) requires that one of the quark bilinears $(\overline{Q} \, u_R)$ and $(\overline{d_R} \, Q)$ or both be $\mathcal{G}$ non-singlets. The latter implies that the Yukawa couplings of $u$ and $d$ quarks
\begin{equation}
\label{eq:Yukawa-ud}
H^{\dagger} \, \overline{Q} \, u_{R}, \quad H \, \overline{Q} \, d_{R}
\end{equation}
are not $\mathcal{G}$-invariant and forbidden by this symmetry. Therefore, the light quarks do not receive their masses as a result of the EWSB. In principle, this is in line with the fact that the light quarks $u, d$ are particular among other quarks by being much lighter than the others. However, the statement of vanishing masses $m_{u, d} = 0$, or even one of them, 
seems to contradict the known results of lattice calculations \cite{Allton:2008pn} and experimental data on the light meson masses.
Therefore, small $m_{u, d} \ne 0$ must be generated in some way to make our scenario viable.
In principle, for this scenario it is not necessary that both Yukawa couplings in Eq.~(\ref{eq:Yukawa-ud}) are forbidden. As seen from (\ref{eq:Eff-Oper-dim-7-1}), it is sufficient that only one of them, say the $u$-quark Yukawa coupling, be forbidden as suggested in Ref.~\cite{Tho92}.  

Thus, we assume that $m_{d}$ is generated via an effective Yukawa coupling (\ref{eq:Yukawa-ud}),  realized at some loop level for making it sufficiently small in comparison to the other heavier quarks. On the other hand, we require that the Yukawa coupling for u-quark be forbidden by the symmetry $\mathcal{G}$ so that above the electroweak scale the current u-quark mass is $m_{u}=0$.
Therefore, in this setup we require that the quark bilinears transform under  the $\mathcal{G}$-symmetry group as 
\begin{eqnarray}\label{eq:bilin-m0-1}
\mathcal{G}-\mbox{non-singlet}: (\overline{Q} \, u_R), \ \ \ \  
\mathcal{G}-\mbox{singlet}: (\overline{d_R} \, Q). 
\end{eqnarray}
Consequently, we should set $g^{d}=0$ in Eq.~(\ref{eq:Eff-Oper-dim-7-1}).
Note that the case of the vanishing current $u$-quark mass has long been considered as one of the possible solutions of the strong CP problem, allowing to rotate away the CP-violating angle $\theta$ from the QCD Lagrangian.  

However, the key question here is whether this is compatible with the lattice value 
$m^{latt}_{u}=2.78\pm 0.19 $ MeV  \cite{Allton:2008pn} and the light meson masses.
We start with the observation that $m_{u}=0$ at some high-energy cutoff scale 
does not prevent generation of a non-zero effective quark mass $m^{eff}_{u}$ at low sub-GeV scales. 
%
Here there are several sources of $m^{eff}_{u}\neq 0$ rooting in the strong interaction dynamics. 




First, we note that in a generic effective theory the light quark masses can be generated due to chiral symmetry breaking via the effective SM-invariant operators
\begin{equation}
\label{eq:Quark-operators-1}
%
%
\mathcal{O}_6^{qq} = \frac{\kappa^{qq}}{\Lambda_{qq}^2} \, \overline{Q} \, Q_R \, \overline{Q_R} \, Q, \quad
%
%
\mathcal{O}_6^{ud} = \frac{\kappa^{ud}}{\Lambda_{ud}^2} \, \overline{Q} \, u_R \, \overline{Q} \, d_R,
\end{equation}
with 
$Q^{T}_{R} = (u,d)_{R}$ being right-handed isodublet. Here, $\Lambda_{qq}, \Lambda_{ud}$ are scales of the physics underlying these operators. 
Both operators in Eq.~(\ref{eq:Quark-operators-1}) among their components have
\begin{align}
\label{eq:mass-components-1}
& \mathcal{O}_6^{qq} \sim \overline{u_L} \, u_R \, \overline{u_R} \, u_{L} + \overline{d_L} \, d_R \, \overline{d_R} \, d_L, \nonumber \\
& \mathcal{O}_6^{ud} \sim \overline{u_L} \, u_R \, \overline{d_L} \, d_R,
\end{align}
which can contribute to the effective light-quark mass $m^{eff}_{u, d}$ after spontaneous breaking of chiral symmetry and formation of the quark condensate $\langle \bar{q}q\rangle$. We note that the operator $\mathcal{O}_6^{qq}$ conserves chiral symmetry while 
$\mathcal{O}_6^{ud}$ breaks it explicitly. In our setup (\ref{eq:bilin-m0-1}) the operator 
$\mathcal{O}_6^{ud}$ is forbidden by $\mathcal{G}$-symmetry. In a scenario with 
\mbox{$m_{u}=m_{d}=0$} above the electroweak scale this operator is allowed and can have interesting implications, if its scale $\Lambda_{ud}$ is not very high. This scenario will be addressed elsewhere.

The chiral symmetric operator $\mathcal{O}_6^{qq}$  in Eq.~(\ref{eq:Quark-operators-1}) is well known in the context of the Nambu--Jona-Lasinio model considered as chiral low-energy effective theory of QCD.
%
Recall, that in this approach the one-gluon exchange diagram with the amplitude
\begin{equation}
\label{eq:NJL-1}
\left( \overline{Q} \, \gamma^{\mu} \, \lambda^a \,Q   \right) D^{(G)ab}_{\mu \nu} \left( \overline{Q_R} \, \gamma^{\nu} \, \lambda^b \, Q_R\right)
%
\end{equation}
turns to a point-like 4-quark operator in a truncated theory, where
the gluon propagator $D^{(G)}(k^2)$ is replaced with 
$g^{\mu \nu}/\Lambda^{2}_{QCD}$.
Here, $\Lambda_{QCD}\sim 100$ MeV is a characteristic scale of non-perturbative QCD. After Fierz rearrangement in (\ref{eq:NJL-1}) one finds the operator $\mathcal{O}_6^{qq}$  in Eq.~(\ref{eq:Quark-operators-1}) with the scale 
$\kappa^{qq}/ \Lambda^{2}_{qq} \sim -\alpha_s/(4 \Lambda_{QCD}^2)$.
After spontaneous breaking of chiral symmetry this operator renders a contribution 
%
\begin{equation}
\label{eq:UD-masses-ChBr} 
m^{eff}_{u,d} = m^{C}_{u, d} = \kappa \, \frac{\langle \overline{q} q \rangle}{\Lambda_{qq}^2} = \frac{\alpha_{s}}{4} \, \omega \left( \frac{\omega}{\Lambda_{QCD}} \right)^2\sim \omega
\end{equation}
to the masses of  $u$ and $d$ quarks, converting them to the so-called constituents quarks with an effective mass \mbox{$m^{C}_{q}\sim 100$~MeV}. 


However, spontaneous breaking of chiral symmetry cannot be the only source of the quark masses.  
They must also have a piece $m_{u,d}\neq 0$, which breaks chiral symmetry explicitly.  According to the Gell-Mann--Oakes--Renner relation (\ref{eq:GMOR-1}), this is needed in order to pions, as Goldstone bosons of the spontaneous breaking of  chiral symmetry, acquire non-zero masses. 
In our setup (\ref{eq:bilin-m0-1})  \mbox{$d$-quark} has $m_{d}\neq 0$ at a high-energy cutoff scale due to Yukawa coupling (\ref{eq:Yukawa-ud}) explicitly breaking chiral symmetry. 
However, $u$-quark is also  required to contribute to this explicit breaking as follows from the analysis of  the meson mass spectrum (see, for instance Ref.~\cite{Kitazawa:2017hqk} and references therein).  Here we adopted $m_{u}= 0$ 
above the electroweak scale.  
%
%
In Ref.~\cite{Davoudiasl:2005ai} it was advocated that the next to leading order chiral Lagrangian terms together with the QCD instanton  are able to induce at the QCD scale $\Lambda_{\rm QCD}\sim 100$ MeV a contribution to 
$u$-quark mass explicitly breaking chiral symmetry.  
The resulting effective mass is compatible with the lattice result \cite{Allton:2008pn}. The value of the $u$-quark effective mass due to these two was estimated in Ref.~\cite{Kitazawa:2017hqk} with the result
\begin{eqnarray}\label{eq:ChSB-mass}
m^{ChSB}_{u}= 2.33\pm 0.20 \mbox{MeV}. 
\end{eqnarray}
According to \cite{Kitazawa:2017hqk} this value is compatible with the light meson masses. However, there is certain tension with the lattice value $m^{latt}_{u}=2.78\pm 0.19 $ MeV  \cite{Allton:2008pn}. In our opinion this situation requires further study and clarification. 
Having this point in mind we adopt in this paper the setup  (\ref{eq:bilin-m0-1}) and examine its phenomenological consequences.

\begin{figure}[t]
\vspace{-40mm}
\includegraphics[width=\columnwidth]{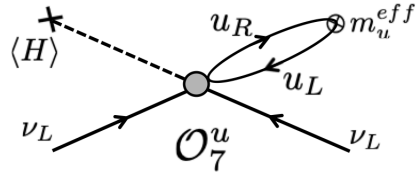}
\vspace{-40mm}
\caption{The loop contribution, $m^{1-loop}_{\nu}$, of  the operator $\mathcal{O}_7^{u}$ to the neutrino mass.
}
\label{fig:Loop-Nu} 
\end{figure}

Before this, the following important comment is in order.  In our scenario it is crucial that a non-zero effective mass of \mbox{$u$-quark}, $m^{eff}_{u}$, is generated at a low scale of the order of the typical QCD scale $\Lambda_{QCD}\sim 100$ MeV.  In fact, if \mbox{$u$-quark} acquires a mass $m^{eff}_{u}\neq 0$, regardless of its origin, one can close the $\bar{Q}u_{R}$ legs of the operator  
$\mathcal{O}_7^{u}$ in Eq.~(\ref{eq:Eff-Oper-dim-7-1}) via the $u$-quark mass term  
\footnote{We are thankful to Martin Hirsch for drawing our attention to this fact.}
 $m^{eff}_{u}$ as shown in Fig.~\ref{fig:Loop-Nu}. 
This will lead to a 1-loop contribution to the Majorana neutrino mass, which can be estimated as
\begin{eqnarray}\label{eq:u-loop-NuMass-1}
m^{1-loop}_{\nu} &\sim& \frac{1}{\sqrt{2}} \frac{g^{u}_{\alpha\beta}}{4 \pi^{2}} v\frac{1}{\Lambda^{3}} 
m^{eff}_{u} \Lambda^{2}_{u},
\end{eqnarray}
where, $\Lambda_{u}$ is a scale at which $m^{eff}_{u}$ is generated. 
In our model $m^{eff}_{u} =  m^{C}_{u} +m^{ChSB}_{u}$, where 
$m^{C}_{u}$ and $m^{ChSB}_{u}$ are given in 
(\ref{eq:UD-masses-ChBr}) and (\ref{eq:ChSB-mass}), respectively. Both these contributions are generated at a scale around $\Lambda_{QCD}\sim 100$~MeV. At a higher scales they are rapidly decreasing, as any non-perturbative QCD effect, providing cutoff in the loop integral in Fig.~\ref{fig:Loop-Nu}. Thus, we substitute $\Lambda_{u}=\Lambda_{QCD}$, $m^{eff}_{u} =  m^{C}_{u} +m^{ChSB}_{u}\approx m^{C}_{u}$ in 
(\ref{eq:u-loop-NuMass-1}) and have the following neutrino mass with the considered 1-loop correction
\begin{eqnarray}\label{eq:u-loop-NuMass-2}
m_{\alpha \beta}^{\nu} \simeq
\frac{g_{\alpha \beta}}{\sqrt{2}} \, v \left( \frac{\omega}{\Lambda} \right)^3\left(1 + \frac{\alpha_{s}}{16\pi^{2}} \right). 
\end{eqnarray}
The 1-loop correction is small and irrelevant for our estimations based on Eq.~(\ref{eq:Dim-7-Mnu}) with $g_{\alpha\beta} = g^{u}_{\alpha\beta}$. Let us recall that in our setup (\ref{eq:bilin-m0-1}) we have $g^{d}_{\alpha\beta }= 0$.  It is worth noting that in general the loop in Fig.~\ref{fig:Loop-Nu} could present a problem. In the case, if  
\mbox{$\Lambda_{u}\sim \Lambda \sim 1$ TeV}  its contribution to neutrino mass would be unacceptably large $m_{\nu} \sim 10^{3}-10^{5}$ eV.  As we have shown our setup is free of this problem.
%
%

\section{Limits on LNV Lepton-Quark Interactions}
\label{sec:lim-eps}
Let us examine phenomenological limits on the strength of the effective LNV lepton-quark interactions predicted by the QCSS. These interactions are derived from (\ref{eq:Eff-Oper-dim-7-1}) and (\ref{eq:neutrino-quark coupling}). It is convenient to 
introduce the dimensionless parameters
\begin{eqnarray}
\label{eq:DimLessPar-1-1}
\varepsilon_{\alpha \beta} & =& \frac{g_{\alpha \beta} \, v/\Lambda^3}{G_{\mathrm{F}}},
\end{eqnarray}
giving a measure of the relative strength of the four-fermion interactions (\ref{eq:neutrino-quark coupling}) with respect to the Fermi constant $G_{\mathrm{F}} \approx 1.166 \times 10^{-5} \, \mathrm{GeV}^{-2}$ of the standard weak interactions. We denoted $g_{\alpha\beta} \equiv g_{\alpha\beta}^{u}$. The latter was introduced in Eq.~(\ref{eq:Eff-Oper-dim-7-1}). As discussed in the previous section, in our setup (\ref{eq:bilin-m0-1}) we set $g_{\alpha\beta}^{d}=0$.

\begin{figure}[t]
\centering
\includegraphics[width=\columnwidth]{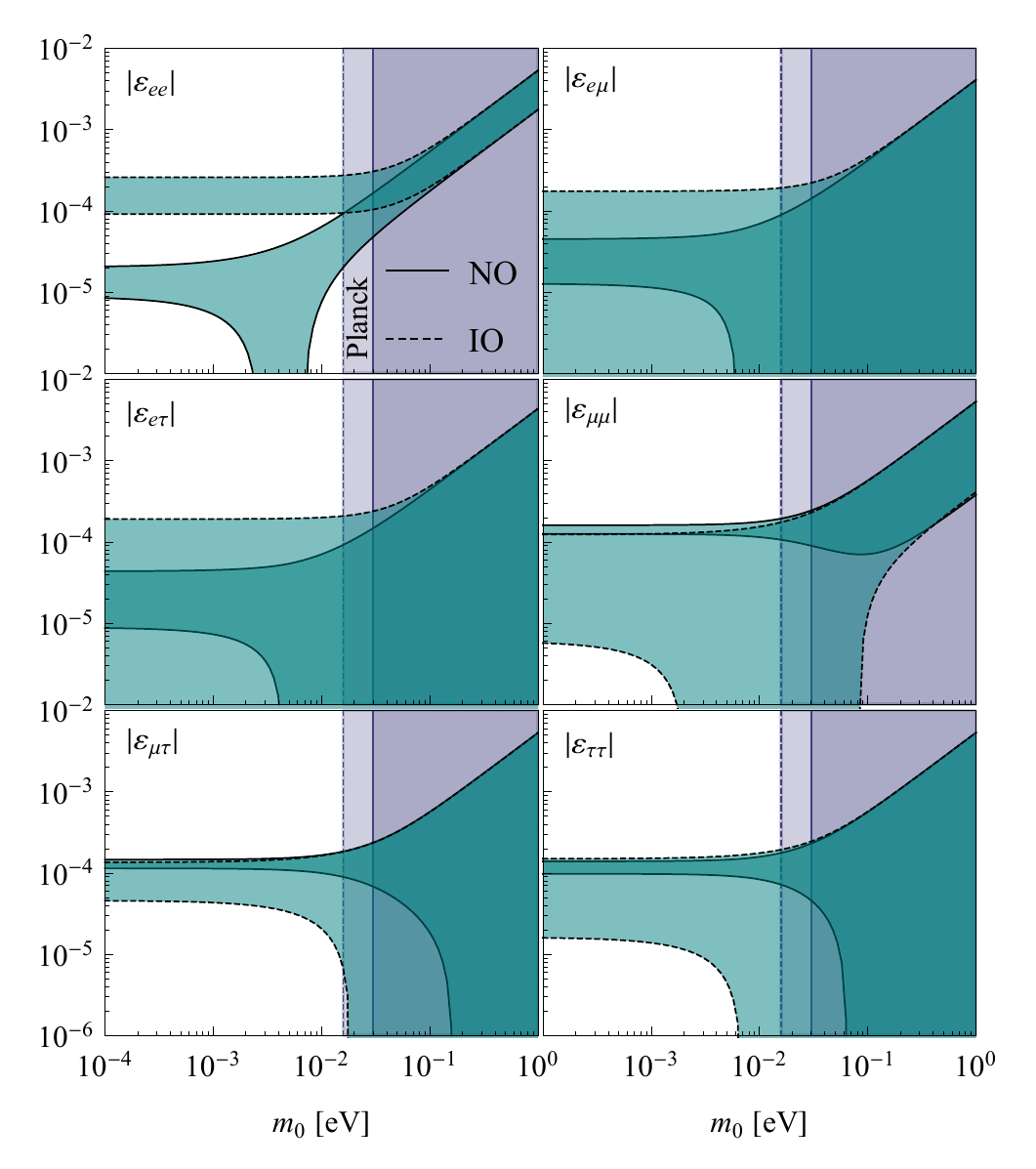}
\caption{\label{fig-2}Predictions of QCSS for the nonstandard neutrino-quark couplings $|\varepsilon_{\alpha \beta}|$ defined in Eq.~(\ref{eq:DimLessPar-12}) vs.\ the lightest-neutrino mass $m_0$.
%
The allowed gray regions between the curved lines are derived using current best-fit values of the neutrino-oscillation parameters \cite{deSalas:2017kay} and CP phases $\delta \in [0, \, 2 \pi)$ and $\alpha_1, \, \alpha_2 \in [0, \, \pi)$.
The solid and dashed lines refer to the normal (NO) and inverted (IO) orderings of the neutrino masses, respectively. 
The vertical gray bands represent the regions excluded by 
\textit{Planck} measurements at $95\% \, \mathrm{C.L.}$ \cite{Agh18}.
The horizontal gray bands in the first plot correspond to the regions excluded by KamLAND-Zen $90\% \, \mathrm{C.L.}$ limits on $0 \nu \beta \beta$-decay \cite{Gan16}. The solid and dashed horizontal limiting lines refer to the scenarios with and without the nuclear-matter effect discussed in Section~\ref{sec:NLDBD}, respectively.}
\end{figure}

Here, assuming the dominance of the QCSS in the Majorana mass of neutrinos, we can extract limits on $\varepsilon_{\alpha \beta}$ from the neutrino-oscillation data since, according to Eq.~(\ref{eq:Dim-7-Mnu}), they are directly related to the elements of the neutrino mass matrix 
\begin{equation}
\label{eq:DimLessPar-12}
\varepsilon_{\alpha \beta} = \frac{g_{\alpha \beta} \, v/\Lambda^3}{G_{\mathrm{F}}} = -\frac{m_{\alpha \beta}^{\nu}/\langle \overline{q} q \rangle}{G_{\mathrm{F}}/\sqrt{2}}.
\end{equation}
Using Eqs.~(\ref{eq:e-e})--(\ref{eq:tau-tau}), we relate these LNV lepton-quark parameters to the neutrino-oscillation parameters. The current values of the latter we take from Ref.~\cite{deSalas:2017kay}.
Then, varying
the CP phases in the intervals $\delta \in [0, \, 2 \pi)$ and $\alpha_1, \, \alpha_2 \in [0, \, \pi)$, we find the exclusion plots in the plane $(m_0, \, |\varepsilon_{\alpha \beta}|)$ shown in Fig.~\ref{fig-2} for the best-fit values of the neutrino oscillation parameters $\theta_{ij}$ and $\Delta m_{ij}^2$ ($i, \, j = 1, \, 2, \, 3$).

As is known, the upper cosmological limit on the sum of neutrino masses
\begin{equation}
\label{eq:Planck}
\sum_i m_i < 0.12 \, \mathrm{eV} \textnormal{ at } 95\% \, \mathrm{C.L.},
\end{equation}
set by the \textit{Planck} measurements \cite{Agh18,Vagno17}, impose the limit on the mass $m_0$ of the lightest neutrino. Applying Eqs.~(\ref{eq:NO-m})--(\ref{eq:IO-m}), one finds
\begin{equation}
\label{eq:Planck-m0}
m_0 < 30.1 \, \mathrm{meV}, \, 15.9 \, \mathrm{meV}
\end{equation}
for the normal (NO) and inverted (IO) neutrino-mass orderings, respectively.
These limits are shown in Fig.~\ref{fig-2} as vertical bands.

It is instructive to show allowed ranges $(\varepsilon_{\alpha \beta}^{\mathrm{min}}; \, \varepsilon_{\alpha \beta}^{\mathrm{max}})$ of the LNV lepton-quark interaction parameters $\varepsilon_{\alpha \beta}$ from (\ref{eq:DimLessPar-12}). These ranges extracted from the exclusion plots in Fig.~\ref{fig-2} for NO and IO are
\begin{align}
\label{eq:eps-NO-1}
|\varepsilon_{\alpha \beta}^{\mathrm{NH}}| & =
\begin{pmatrix}
(0; \, 1.7) & (0; \, 1.3) & (0; \, 1.5) \\
& (0.9; \, 2.4) & (0.7; \, 2.4) \\
& & (0.5; \, 2.3)
\end{pmatrix} \times 10^{-4}, \\
\label{eq:eps-IO-1}
|\varepsilon_{\alpha \beta}^{\mathrm{IH}}| & =
\begin{pmatrix}
(0.9; \, 2.7) & (0; \, 1.9) & (0; \, 2.1) \\
& (0; \, 1.7) & (0.1; \, 1.8) \\
& & (0; \, 1.9)
\end{pmatrix} \times 10^{-4}.
\end{align}

To the best of our knowledge, the only analysis of phenomenological limits on the lepton-quark interaction strength $\varepsilon$ from Eq.~(\ref{eq:DimLessPar-1-1}) existing in the literature is given in Refs.~\cite{Tho92,Kovalenko:2013eba} where the SN~1987A and meson decays were studied. 
In the former case 
these limits are in the range $\varepsilon < 10^{-3}$, which is an order of magnitude weaker than our limits in Eqs.~(\ref{eq:eps-NO-1})--(\ref{eq:eps-IO-1}). As to the LFV meson decays, reasonable limits on \mbox{$\varepsilon$-parameters} cannot be extracted from the experimental data. Indeed, considering as an example the LNV decay 
\mbox{$K^{+}\rightarrow \pi^{-} \mu^{+}\mu^{+}$}, one finds \cite{Helo:2010cw} for its branching ratio \mbox{Br$(K^{+}\rightarrow \pi^{-} \mu^{+}\mu^{+})\sim |\varepsilon_{\mu\mu}|^{2}\times 10^{-30}$}, which should be compared with  the current experimental  upper bound \mbox{Br$(K^{+}\rightarrow \pi^{-} \mu^{+}\mu^{+}) \leq 10^{-11}$}. Of course, this gives no practical information on the $\varepsilon_{\mu\mu}$-parameter.

\section{Quark Condensate Seesaw in Neutrinoless Double-Beta Decay}
\label{sec:NLDBD}
\begin{figure}[t]
\centering
\includegraphics[width=\columnwidth]{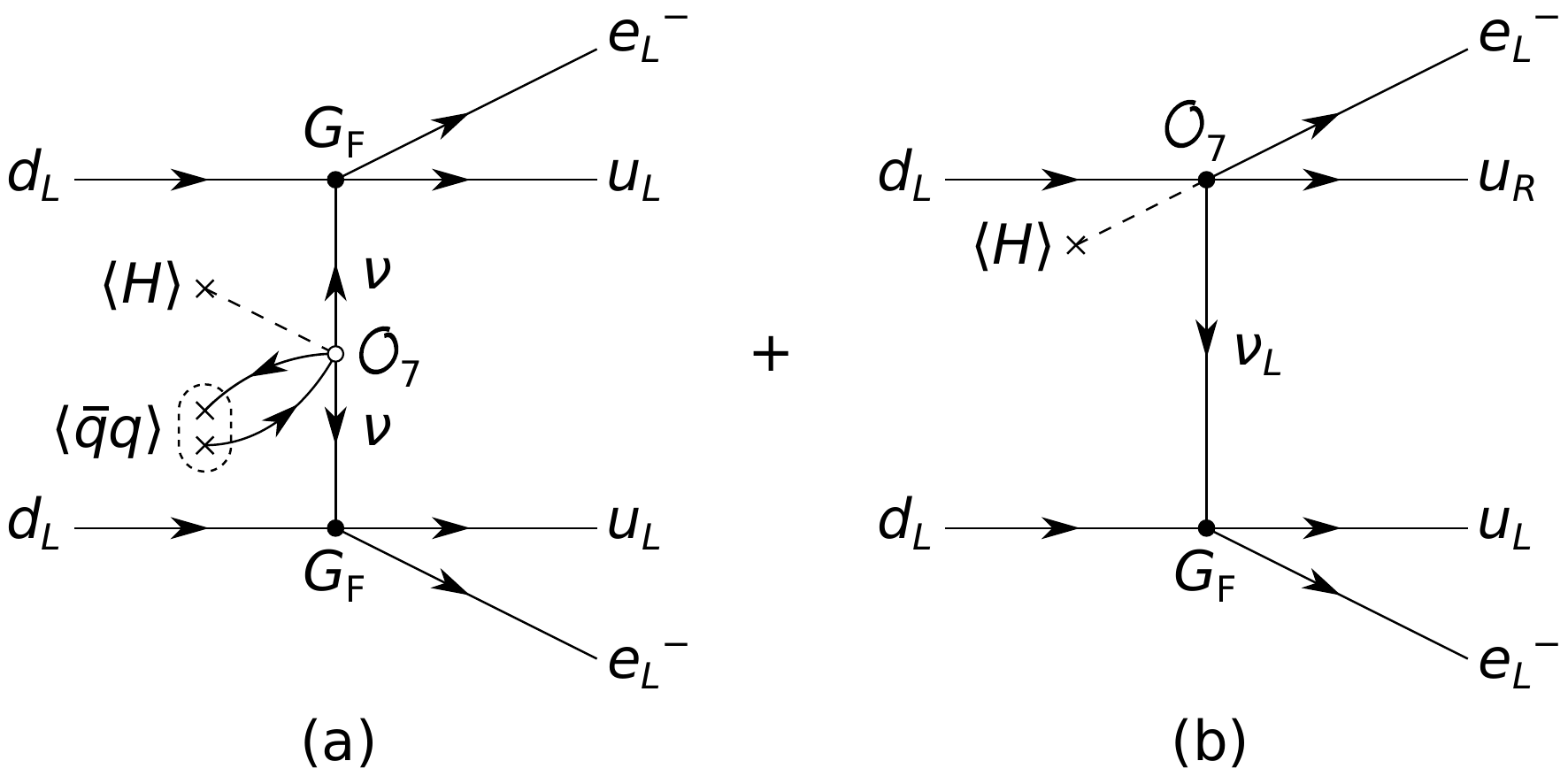}
\caption{\label{fig:O7-NLDBD}Contributions of the effective operators (\ref{eq:Eff-Oper-dim-7-1}) to the $0 \nu \beta \beta$ decay.}
\end{figure}

Let us consider the contribution of the operator $\mathcal{O}_7^{u}$ 
in Eq.~(\ref{eq:Eff-Oper-dim-7-1}) to $0 \nu \beta \beta$ decay.
After the EWSB this operator generates the following interactions relevant to 
\mbox{$0 \nu \beta \beta$-decay} 
\begin{eqnarray}
\label{eq:d7-e-nu-ud}
\mathcal{L}_7 & =& \frac{G_{\mathrm{F}}}{\sqrt{2}} \,
\varepsilon_{ee}\ \left(\overline{e_L} \, \nu_L^C  \, \overline{u_R} \, d_L  
 +
\overline{\nu_L^C} \, \nu_L \overline{u_R} \, u_L\right)  + \mathrm{H.c.},\hspace{9mm}
\end{eqnarray}
where $\varepsilon_{ee}$ is defined in Eq.~(\ref{eq:DimLessPar-1-1}). Let us examine the contribution of the effective $\Delta L = 2$ interaction terms (\ref{eq:d7-e-nu-ud}) to $0 \nu \beta \beta$-decay. The first term (\ref{eq:d7-e-nu-ud}) combined with the SM weak charged-current interaction leads to the contribution shown in Fig.~\ref{fig:O7-NLDBD}(b), which is independent of neutrino mass in the propagator due to chiralities in the vertices $P_L \, (m_{\nu} + \slashed{q}) \, P_R = \slashed{q}$. This is a manifestation of the fact that the $\Delta L = 2$ is not provided by $m_{\nu}$, but solely by the upper vertex in Fig.~\ref{fig:O7-NLDBD}(b).
In our QCSS model the second term in Eq.~(\ref{eq:d7-e-nu-ud}) also contributes to the $0 \nu \beta \beta$-decay via the neutrino-mass mechanism shown in Fig.~\ref{fig:O7-NLDBD}(a). This happens due to the chiral symmetry breaking and formation of the quark condensate. As we discussed in the previous sections, this term is the only source of the neutrino mass in the present model. However, there is a subtlety with the diagram in Fig.~\ref{fig:O7-NLDBD}(a). It describes a process taking place in the nuclear environment, where the chiral quark condensate $\langle \overline{q} q \rangle_N$ is suppressed with respect to the one in the vacuum $\langle \overline{q} q \rangle$. As briefly discussed in Appendix~\ref{sec:Quark Condensate}, this suppression is estimated to be a factor-two effect (\ref{eq:cond-in-nucl}).
%
In what follows, we take this fact into account.

The inverse $0 \nu \beta \beta$-decay half-life in the QCSS model reads (see Appendix~\ref{sec:NME})
\begin{align}
\label{eq:Half-Life-QCSS}
\left( T_{1/2}^{0 \nu} \right)^{-1} & = \left| \varepsilon_{ee} + \frac{m_{\beta \beta}}{m_e} f_{\rm nme} \right|^2\,g^2_A\,
|{M}^{\epsilon}|^2 \, G^{0 \nu} \, \nonumber \\
& = |\varepsilon_{ee}|^{2}\,  \left| 1 + a_\nu f_{\rm nme} \right|^2\, g^2_A \, |{M}^{\epsilon}|^2  \, G^{0 \nu} \, 
\end{align}
where  $m_e$ is the mass of electron. 
The standard kinematical phase-space factor $G^{0 \nu}$ is given in Table~\ref{tab:nme}
and the nuclear structure factor $f_{\rm nme}$ is given by
\begin{equation}
\label{eq:fnme}
f_{\rm nme} = \frac{g_A~ {M}^{\nu}}{{M}^{\varepsilon}}.
\end{equation}
Nuclear matrix elements ${M}^{\nu}$ and ${M}^{\varepsilon}$ associated with mechanisms in Fig. \ref{fig:O7-NLDBD}~(a) and (b),
respectively, are presented in Appendix~\ref{sec:NME}.
$g_A \simeq 1.27$ is the axial-vector weak nucleon coupling constant.

In the second row of Eq.~(\ref{eq:Half-Life-QCSS}) we used the prediction of the QCSS model
\begin{equation}
\label{eq:bb-QCSS-1}
m_{\beta \beta} = a_\nu \, m_e \, \varepsilon_{ee}, \quad a_\nu = \frac{\langle \overline{q} q \rangle_N}{\sqrt{2} \, m_e} \, G_{\mathrm{F}}
=-1.83 \times 10^{-4}.
\end{equation}
Here, $\langle \overline{q} q \rangle_N$ is the value of quark condensate in the nuclear environment of a decaying nucleus, which is different from the vacuum value 
$\langle \overline{q} q \rangle$ as displayed in 
Eq.~(\ref{eq:cond-in-nucl}).
Consequently, in the QCSS we have 
\begin{equation}
\label{eq:mbb-mee-rel}
m_{\beta \beta} = \frac{\langle \overline{q} q \rangle_N}{\langle \overline{q} q \rangle} \, m_{ee}^{\nu} \approx 0.5 \, m_{ee}^{\nu},
\end{equation}
where $m_{ee}^{\nu}$ is given by (\ref{eq:Dim-7-Mnu}) with the chiral condensate in the vacuum.  This result contrasts with the conventional neutrino mass models, where 
$m_{\beta \beta}=m_{ee}^{\nu}$ in the diagonal charged-lepton basis.

With the values of NME for $\ce{^{136}Xe}$ given in Appendix~\ref{sec:NME}, quark condensate in matter (\ref{eq:cond-in-nucl}) with (\ref{eq:quark-condensate-value}) and other known parameters we find 
\begin{equation}
\label{eq:a-b-estimates}
a_\nu ~f_{\rm nme} \simeq -2.00 \times 10^{-7}.
\end{equation}
Recall that, the contributions to $0\nu\beta\beta$-decay amplitude of the diagrams in Fig.~\ref{fig:O7-NLDBD}(a) and Fig.~\ref{fig:O7-NLDBD}(b)
are proportional to the first and second terms in vertical brackets of Eq. (\ref{eq:Half-Life-QCSS}), respectively.
Thus, due to (\ref{eq:a-b-estimates}) the diagram Fig.~\ref{fig:O7-NLDBD}(b) dominates in our scenario.

From the currently most stringent upper bound on the $0 \nu \beta \beta$-decay half-life obtained for $\ce{^{136}Xe}$ by the KamLAND-Zen experiment \cite{Gan16}
\begin{equation}
\label{eq:KamLAND-Zen-1}
T_{1/2}^{0 \nu} > 1.07 \times 10^{26} \, \mathrm{yr} \textnormal{ at } 90\% \, \mathrm{C.L.},
\end{equation}
we find, using (\ref{eq:Half-Life-QCSS}), an upper bound
\begin{equation}
\label{eq:Lim-12}
 \left|\varepsilon_{ee}\right| < 2.49 \times 10^{-10}.
\end{equation}

Comparing the limit (\ref{eq:Lim-12}) for $\varepsilon_{ee}$ with the excluded regions in the first plot of Fig.~\ref{fig-2}, derived from the neutrino-oscillation data, we conclude that the QCSS predicts the NO and 
a rather narrow interval of the lightest neutrino mass
\begin{equation}
\label{eq:m0-range}
2.65 \, \mathrm{meV} < m_0 = m_1 < 6.84 \, \mathrm{meV}.
\end{equation}
Using the $1 \sigma$ ranges of the neutrino-oscillation parameters $\sin^2 \theta_{ij}$ and $\Delta m_{ij}^2$ from Ref.~\cite{deSalas:2017kay}, we derive according to  Eq.~(\ref{eq:NO-m})
the following ranges for the other two neutrino masses
\begin{align}
\label{eq:m0-range-1}
9.0 \, \mathrm{meV} & < m_2 < 11.2 \, \mathrm{meV}, \\
\label{eq:m0-range-2}
49.8 \, \mathrm{meV} & < m_3 < 50.8 \, \mathrm{meV}.
\end{align}   
From the limits (\ref{eq:m0-range})-(\ref{eq:m0-range-1}) we find
the corresponding range for the cosmological neutrino parameter
\begin{equation}
\label{eq:lim-cosm-QCSS-1}
61.4 \, \mathrm{meV} < \Sigma = \sum_i m_i < 68.8 \, \mathrm{meV}.
\end{equation}
This prediction of the QCSS is within the Planck limit (\ref{eq:Planck}).
We also derive the QCSS range for the single-beta-decay parameter
\begin{equation}
\label{eq:Single-Beta-QCSS-1}
9.0 \, \mathrm{meV} < m_{\beta} = \sqrt{\sum_i |U_{ei}|^2 \, m_i^2} < 11.4 \, \mathrm{meV},
\end{equation}
which is beyond the reach of the current and near future tritium beta-decay experiments  (for a recent review see, for instance, Ref.~\cite{Drexlin:2013lha,Aker:2019uuj}).

At last, the limit (\ref{eq:Lim-12}) translated into the $0 \nu \beta \beta$-decay parameter gives the upper bound
\begin{equation}
\label{eq:NLDBB-lim-QCSS-1} 
|m_{\beta \beta}| < 2.33 \times 10^{-5} \, \mathrm{meV}.
\end{equation}
It is worth recalling that in the QCSS this parameter characterizes the subdominant contribution to the $0 \nu \beta \beta$ decay shown in Fig.~\ref{fig:O7-NLDBD}(a), while the dominant one is given by the diagram Fig.~\ref{fig:O7-NLDBD}(b).

\section{Particular Realization of QCSS Model}
\label{sec:Particular realization of QCSS-model}
There is one potential flaw in the model described above: the quark bilinear $(\overline{d_R} \, Q)$, being a $\mathcal{G}$-singlet, allows the $d$-quark tree-level Yukawa coupling shown in (\ref{eq:Yukawa-ud}).
It leads to a tree-level $d$-quark mass $m_{d}$ after the EWSB, which makes its the smallness rather weird.  
The common wisdom, allowing to avoid a fine-tuning,  is to impose on theory an additional softly-broken symmetry forbidding the tree-level Yukawa couplings of the light quarks, but allowing them at certain loop level (for a recent review see, for instance Ref.~\cite{Cai:2017jrq}). 
To this end we can extend the previously used group  $\mathcal{G}$ to $\mathcal{G}^{\prime} = \mathcal{G}\times \mathcal{G}^{d}$ requiring that all the fields, except for $d$-quark, be neutral with respect to the subgroup $\mathcal{G}^{d}$. In this way we can 
forbid with the help of $\mathcal{G}^{d}$ the tree-level d-quark Yukawa coupling. Once this this symmetry is softly broken, the $d$-quark Yukawa can appear at loop level.
In this case the d-quark mass could gain the necessary loop suppression. Let us give an a example of such symmetry group $\mathcal{G}^{\prime}$ of our model
\begin{eqnarray}\label{eq:G-example}
\mathcal{G}^{\prime} = \mathbb{Z}_4\times \mathbb{Z}_{2} \ \xrightarrow[]{soft}
\mathbb{Z}_{4} \, 
\end{eqnarray} 
with the  following  $\mathbb{Z}_4\times \mathbb{Z}_{2}$ - assignment of the fields
\begin{eqnarray}\label{eq:assignment-1}
H &\sim& (1,1), \ \ \ \, Q \sim (1,1), \ \  u_R \sim (-1,1),\\ 
\nonumber 
d_R &\sim& (1,-1),  \ \  L \sim (i,1), \ \  e_R \sim (i,-1).
\end{eqnarray}
Here, we limit ourselves only to the first generation of the fermions. 
With this field assignment the Yukawa couplings of  $u$ and $d$-quarks (\ref{eq:Yukawa-ud}), the analogous electron coupling $H \, \overline{L} \, e_R$, as well as the operators 
$\mathcal{O}_W$ in Eq.~(\ref{eq:Weinberg-1}), $\mathcal{O}_7^d$ in Eq.~(\ref{eq:Eff-Oper-dim-7-1}) and $\mathcal{O}_6^{ud}$ in Eq.~(\ref{eq:Quark-operators-1}) are all forbidden by 
$\mathbb{Z}_{4}\times\mathbb{Z}_{2}$ group.  On the other hand, this group allows the operators $\mathcal{O}_7^u$ in Eq.~(\ref{eq:Eff-Oper-dim-7-1}) and  $\mathcal{O}_6^{qq}$ in Eq.~(\ref{eq:Quark-operators-1}).
Soft symmetry breaking (\ref{eq:G-example}) let the $d$-quark and electron Yukawa couplings arise at certain loop level. As a result these couplings acquire   
loop suppression factors necessary to make $m_{d}$ and $m_{e}$ smaller in comparison to the other SM fermions. 
The loop order depends on the concrete UV model.
In principle, we can introduce an extra loop suppression to the electron Yukawa couplings in order to achieve $m_{d} >m_{e}$. This can be easily done by the extension of the scenario (\ref{eq:G-example}) to
\begin{eqnarray}\label{eq:example-2}
\mathcal{G}^{\prime\prime} = \mathbb{Z}_4\times \mathbb{Z}_{2}\times \mathbb{Z}^{e}_{2} \ \xrightarrow[]{soft[1]}
\mathbb{Z}_{4}\times \mathbb{Z}^{e}_{2}\xrightarrow[]{soft[2]}\mathbb{Z}_{4}
\end{eqnarray}
with all the field neutral with respect to $\mathbb{Z}^{e}_{2}$, but electron having  $(-1)$ assignment to this subgroup. In this case the electron Yukawa appears at the second stage of the soft symmetry breaking chain and, therefore, can be realized at higher loop order than $d$-quark one. In this way the SM fermion mass hierarchy can be generated by sequential loop suppression \cite{CarcamoHernandez:2016pdu} resulting from certain chain of soft symmetry breaking \cite{Arbelaez:2019ofg}.

\section{Conclusion and Discussions}
We studied the Quark Condensate SeeSaw (QCSS) mechanism of generation of the Majorana  neutrino mass matrix due to spontaneous breaking of chiral symmetry. 
The effect of the formation of chiral condensate is transmitted to the neutrino sector via the dimension-$7$ quark-lepton operators $\mathcal{O}_7^{u}$ in 
Eq~ (\ref{eq:Eff-Oper-dim-7-1}). 
They can originate in low-energy limit from a certain class of UV models. 
We imposed on these models 
a symmetry, $\mathcal{G}$, forbidding 
the Weinberg operator $\mathcal{O}_{W}$ in Eq.~(\ref{eq:Weinberg-1}) while allowing the operators $\mathcal{O}_7^{u}$.
In this case the QCSS mechanism dominates over the ordinary tree-level Majorana neutrino mass generated by the electroweak symmetry breaking.  The symmetry $\mathcal{G}$ inevitably forbids the \mbox{$u$-quark} Yukawa coupling making its mass equal to zero at a high-energy cutoff scale.
 We argued, following the existing literature, that $u$-quark receives a non-zero effective mass  
$m^{eff}_{u}$ from non-perturbative QCD effects at the scale \mbox{$\Lambda_{QCD}\sim 100$ MeV}. We commented that $m^{eff}_{u}$ generated in this way is compatible with the light hadron mass spectrum, but shows certain tension with the lattice simulations. 
A detailed study of this issue will be carried out elsewhere.  
We discussed how in this scenario 
$d$-quark and electron can be made naturally lighter than other SM fermions.  We proposed to introduce a softly broken symmetry (\ref{eq:G-example}) or (\ref{eq:example-2}) 
forbidding the tree-level $d$-quark and electron Yukawa couplings, but unlocking them at some loop level. This mechanisms can bring into $d$-quark and electron masses, $m_{d}$ and $m_{e}$,  loop-suppression factors necessary for making them naturally small.
The order of the loop suppression depends on the particular UV model. We postpone the study of the possible UV completions of the QCSS scenario for the subsequent publication.
We also noted that the $u, d$-quark masses always receive a contribution proportional to the chiral condensate via four-quark operators generated by non-perturbative QCD effects, which convert these quarks to the constituent ones.

We derived  the predictions of the QCSS model for the LNV lepton-quark couplings (\ref{eq:DimLessPar-1-1}).
These couplings characterize the nonstandard neutrino-quark and charged lepton-quark interactions arising from the operators in Eq.~(\ref{eq:Eff-Oper-dim-7-1}). They can be relevant for further studies of the phenomenological and astrophysical implications of the QCSS mechanism.

We analyzed the predictions of the QCSS model for neutrinoless double-beta decay. We calculated the corresponding nuclear matrix elements within the Quasiparticle Random Phase Approximation (QRPA) method with partial restoration of the isospin symmetry. 

We commented about the role of 
the nuclear-matter effects for the neutrino mass mechanism in Fig.~\ref{fig:O7-NLDBD}(a).
%
%
We showed that the neutrino mass independent mechanism in Fig.~\ref{fig:O7-NLDBD}(b)  dominates in the QCSS scenario (\ref{eq:bilin-m0-1}).  

We found that the QCSS
predicts the normal ordering (NO) of the neutrino-mass spectrum 
and 
rather narrow ranges (\ref{eq:m0-range})-(\ref{eq:m0-range-1}) for the neutrino masses. This is in accord with the recent global analysis of the neutrino-oscillation data, which favors NO over IO at more than $3\sigma$ \cite{deSalas:2017kay}.  We also derived predictions of the QCSS for some other observables (\ref{eq:lim-cosm-QCSS-1})-(\ref{eq:NLDBB-lim-QCSS-1}). 


\begin{acknowledgments}
This work was supported by the Ministry of Education, Youth and Sports of the Czech Republic under the INAFYM Grant No.~CZ.02.1.01/0.0/0.0/16\_019/0000766 and the Grant of the Plenipotentiary Representative of the Czech Republic in JINR under Contract No.~202 from 24/03/2020, Fondo Nacional de Desarrollo Cient\'{i}fico y Tecnol\'{o}gico (FONDECYT, Chile) No.~1190845 and Agencia Nacional de Investigaci\'{o}n y Desarrollo (ANID, Chile) AFB180002, the RFBR Grant No.~18-02-00733 (Russia), and the VEGA Grant Agency of the Slovak Republic under Contract No.~1/0607/20.
\end{acknowledgments}

\appendix
\section{Majorana Mass Matrix}
\label{sec:Nu-Mass-Matrix}
Diagonalizing the complex symmetric $3 \times 3$ Majorana-neutrino mass matrix $m^{\nu}$ in Eqs.~(\ref{eq:Dim-7-Mnu}) and (\ref{eq:Nu-MassTerm-1}) $U^{\mathrm{T}} m_{\nu} U = \mathrm{diag}(m_1, \, m_2, \, m_3)$ with a unitary
lepton mixing matrix $U$, one gets the usual relation
\begin{equation}
\nu_{\alpha L} = \sum_i U_{\alpha i} \, \nu_{iL}
\end{equation}
between the neutrino mass eigenstates $\nu_i$ with masses $m_i$ and the flavor eigenstates $\nu_{\alpha}$. The matrix $U$ is known as the Pontecorvo--Maki--Nakagawa--Sakata (PMNS) matrix
\begin{align}
U = &
\begin{pmatrix}
1 & 0 & 0 \\
0 & c_{23} & s_{23} \\
0 & -s_{23} & c_{23}
\end{pmatrix}
\begin{pmatrix}
c_{13} & 0 & s_{13} \, e^{-i \delta} \\
0 & 1 & 0 \\
-s_{13} e^{i \delta} & 0 & c_{23}
\end{pmatrix}
\begin{pmatrix}
c_{12} & s_{12} & 0 \\
-s_{12} & c_{12} & 0 \\
0 & 0 & 1
\end{pmatrix} \nonumber \\
& \begin{pmatrix}
e^{i \alpha_1} & 0 & 0 \\
0 & e^{i \alpha_2} & 0 \\
0 & 0 & 1
\end{pmatrix},
\end{align}
%
parameterized in terms of the mixing angles $\theta_{12}$, $\theta_{13}$, $\theta_{23}$ ($s_{ij} \equiv \sin \theta_{ij}$, $c_{ij} \equiv \cos \theta_{ij}$), Dirac phase $\delta$ and Majorana phases $\alpha_1$, $\alpha_2$ \cite{Cap16}.


The neutrino masses $m_i$ ($i = 1, \, 2, \, 3$) can be parameterized by the lightest-neutrino mass $m_0$ (which is unknown) and the mass-squared splittings $\Delta m_{ij}^2 = m_i^2 - m_j^2$
(known from the neutrino-oscillations experiments) for two types of the neutrino-mass ordering as \\
\emph{Normal ordering (NO)} with $m_1 < m_2 \ll m_3$:
\begin{align}
\label{eq:NO-m}
& m_1 = m_0, \quad m_2 = \sqrt{m_0^2 + \Delta m_{21}^2}, \nonumber \\
& m_3 = \sqrt{m_0^2 + \Delta m_{31}^2}.
\end{align}
\emph{Inverted ordering (IO)} with $m_3 \ll m_1 < m_2$:
\begin{align}
\label{eq:IO-m}
& m_1 = \sqrt{m_0^2 - \Delta m_{31}^2}, \nonumber \\
& m_2 = \sqrt{m_0^2 + \Delta m_{21}^2 - \Delta m_{31}^2}, \quad m_3 = m_0.
\end{align}

Elements $m_{\alpha \beta}^{\nu} = m_{\beta \alpha}^{\nu}$ of the Majorana mass matrix depend on the Dirac phase $\delta \in [0, \, 2 \pi)$, Majorana phases $\alpha_1, \, \alpha_2 \in [0, \, \pi)$, and neutrino masses $m_i$ determined by the lightest-neutrino mass $m_0$ and neutrino-mass ordering (NO or IO). Namely
\begin{widetext}
\begin{align}
\label{eq:e-e}
m_{ee}^{\nu} & = c_{12}^2 c_{13}^2 e^{-i 2 \alpha_1} m_1 + s_{12}^2 c_{13}^2 e^{-i 2 \alpha_2} m_2 + s_{13}^2 e^{i 2 \delta} m_3, \\
\label{eq:e-mu}
m_{e \mu}^{\nu} & = -c_{12} c_{13} (s_{12} c_{23} + c_{12} s_{13} s_{23} e^{-i \delta}) e^{-i 2 \alpha_1} m_1 + s_{12} c_{13} (c_{12} c_{23} - s_{12} s_{13} s_{23} e^{-i \delta}) e^{-i 2 \alpha_2} m_2 + s_{13} c_{13} s_{23} e^{i \delta} m_3, \\
\label{eq:e-tau}
m_{e \tau}^{\nu} & = c_{12} c_{13} (s_{12} s_{23} - c_{12} s_{13} c_{23} e^{-i \delta}) e^{-i 2 \alpha_1} m_1 - s_{12} c_{13} (c_{12} s_{23} + s_{12} s_{13} c_{23} e^{-i \delta}) e^{-i 2 \alpha_2} m_2 + s_{13} c_{13} c_{23} e^{i \delta} m_3, \\
\label{eq:mu-mu}
m_{\mu \mu}^{\nu} & = (s_{12} c_{23} + c_{12} s_{13} s_{23} e^{-i \delta})^2 e^{-i 2 \alpha_1} m_1 + (c_{12} c_{23} - s_{12} s_{13} s_{23} e^{-i \delta})^2 e^{-i 2 \alpha_2} m_2 + c_{13}^2 s_{23}^2 m_3, 
\\
\label{eq:mi-tau}
m_{\mu \tau}^{\nu} & = -(s_{12} s_{23} - c_{12} s_{13} c_{23} e^{-i \delta})(s_{12} c_{23} + c_{12} s_{13} s_{23} e^{-i \delta}) e^{-i 2 \alpha_1} m_1 \nonumber \\
& \quad - (c_{12} s_{23} + s_{12} s_{13} c_{23} e^{-i \delta})(c_{12} c_{23} - s_{12} s_{13} s_{23} e^{-i \delta}) e^{-i 2 \alpha_2} m_2 + c_{13}^2 s_{23} c_{23} m_3, \\
\label{eq:tau-tau}
m_{\tau \tau}^{\nu} & = (s_{12} s_{23} - c_{12} s_{13} c_{23} e^{-i \delta})^2 e^{-i 2 \alpha_1} m_1 + (c_{12} s_{23} + s_{12} s_{13} c_{23} e^{-i \delta})^2 e^{-i 2 \alpha_2} m_2 + c_{13}^2 c_{23}^2 m_3.
\end{align}
\end{widetext}
We employ these relations for the analysis of the nonstandard neutrino-quark couplings (\ref{eq:neutrino-quark coupling}) and (\ref{eq:DimLessPar-1-1}), which is done in Section~\ref{sec:lim-eps}. In this analysis, we use the neutrino-oscillation data from Ref.~\cite{deSalas:2017kay}.

%
%

\section{Quark Condensate}
\label{sec:Quark Condensate}
Chiral symmetry is approximate invariance of the QCD Lagrangian under the global $\mathrm{SU}(3)_L \times \mathrm{SU}(3)_R$ transformations in the space of the lightest quark flavors $q = u, \, d, \, s$. Below the chiral scale $4 \pi \, f_{\pi} \sim 1 \, \mathrm{GeV}$, this symmetry is spontaneously broken
by the light-quark condensates $\braket{0 | \overline{q} q | 0} \equiv \langle \overline{q} q \rangle \ne 0$
in the QCD ground state (vacuum) $\ket{0}$. The corresponding Goldstone bosons form the octet of light mesons.
Their nonzero masses originate from the explicit breaking of the chiral symmetry by the light-quark current mass terms in the QCD Lagrangian/Hamiltonian:
\begin{align}
\mathcal{H}_{\mathrm{m}} & = m_u \overline{u} u + m_d \overline{d} d + m_s \overline{s} s \nonumber \\
& = \frac{1}{2} (m_u + m_d) (\overline{u} u + \overline{d} d) \nonumber \\
& \quad + \frac{1}{2} (m_u - m_d) (\overline{u} u - \overline{d} d) + m_s \overline{s} s \nonumber \\
& = 2 m_q \overline{q} q + \dots,
\end{align}
where we separated the isospin-singlet and isospin-triplet quark combinations. In what follows, we retain the singlet and consider only the lightest $u$ and $d$ quarks, denoting $m_q = \frac{1}{2} (m_u + m_d)$ and $\overline{q} q = \frac{1}{2} (\overline{u} u + \overline{d} d)$.
Note that in our scenario the mass $m_u$ of u-quark, explicitly
breaking chiral symmetry, originates not from the electroweak symmetry breaking,
which is the case for all the other current quark masses, but from non-perturbative
QCD. Thus, throughout this section, it is implied that
$m_u = m^{ChSB}_u$, as defined in Eq. (\ref{eq:ChSB-mass}).

Here, we will examine the effect of nuclear environment on the formation of light quark condensate. Following Ref.~\cite{Coh92}, we use the Hellmann--Feynman theorem, allowing one to analyze the condensates in a model-independent way to the first order in nucleon density. The Hellmann--Feynman theorem states
\begin{equation}
\label{eq:HFth-1}
\braket{\psi(\lambda) | \frac{\mathrm{d}}{\mathrm{d} \lambda} H(\lambda) | \psi(\lambda)} = \frac{\mathrm{d}}{\mathrm{d} \lambda} E(\lambda),
\end{equation}
where $\ket{\psi(\lambda)}$ and $E(\lambda)$ are, respectively, the normalized energy eigenstates and eigenvalues of the Hamiltonian $H(\lambda)$ with explicit dependence on the parameter $\lambda$. Choosing $\lambda = m_q$ and $H = \int \mathrm{d}^3 \vec{x} \, \mathcal{H}_{\mathrm{m}}$ we get
\begin{equation}
\label{eq:HFth-2}
2 m_q \braket{\psi(m_q) | \int \mathrm{d}^3 \vec{x} \, \overline{q} q | \psi(m_q)} = m_q \frac{\mathrm{d} E(m_q)}{\mathrm{d} m_q},
\end{equation}
where both parts of this equation are multiplied by $m_q$ to ensure renormalization-group invariance of this relation \cite{Tarrach:1981bi}. Let us consider two cases $\ket{\psi(m_q)} = \ket{0}, \, \ket{\rho_N}$, where $\ket{0}$ is the QCD vacuum and $\ket{\rho_N}$ refers to the ground state of (uniform) nuclear matter at rest with nucleon density $\rho_N$. Writing Eq.~(\ref{eq:HFth-2}) for these two cases, we subtract one from the other and obtain
\begin{equation}
\label{eq:rel-1}
2 m_q \left( \braket{\rho_N | \overline{q} q |\rho_N} - \braket{0 | \overline{q} q | 0} \right) = m_q \frac{\mathrm{d} \mathcal{E}_N}{\mathrm{d} m_q},
\end{equation}
where $\mathcal{E}_N$ is the energy density of nuclear matter. Provided the kinetic and potential energy of nucleons are known to be small, one has
\begin{equation}
\label{eq:rel-12}
\mathcal{E}_N = m_N \rho_N.
\end{equation}
On the other hand, there is a current-algebra relation \cite{Jaffe:1987sw}
\begin{equation}
\label{eq:sigma-1}
\sigma_N = m_q \frac{\mathrm{d} m_N}{\mathrm{d} m_q},
\end{equation}
where $\sigma_N$ is the pion-nucleon sigma term measuring the nucleon-mass $m_N$ shift from the chiral limit $m_{u, d} \to 0$. Then, using Eqs.~(\ref{eq:rel-1}), (\ref{eq:rel-12}) and (\ref{eq:sigma-1}), one finds a model-independent relation \cite{Coh92}
\begin{equation}
\label{eq:rel-123}
\frac{\langle \overline{q} q \rangle_N}{\langle \overline{q} q \rangle} = 1 + \frac{\sigma \rho_N}{2 m_q \langle \overline{q} q \rangle} = 1 - \frac{\rho \sigma_N}{f_{\pi}^2 m_{\pi}^2},
\end{equation}
characterizing the amount of chiral-symmetry restoration in dense medium. Here, we denoted $\langle \overline{q} q \rangle_N \equiv \braket{\rho_N | \overline{q} q |\rho_N}$ and $\langle \overline{q} q \rangle \equiv \braket{0 | \overline{q} q | 0}$. The Gell-Mann--Oakes--Renner relation \cite{Gel68}
\begin{equation}
\label{eq:GMOR-1}
2 m_q \langle \overline{q} q \rangle = -f_{\pi}^2 m_{\pi}^2,
\end{equation}
has been used to derive Eq.~(\ref{eq:rel-123}). In order to estimate the nuclear-matter effect on the quark condensate on the basis of Eq.~(\ref{eq:rel-123}), we adopt the usual value for the nucleon density $\rho = \rho_p + \rho_n = 0.17 \, \mathrm{fm}^{-3}$, the recent large value of $\sigma = 64 \, \mathrm{MeV}$ from a partial-wave analysis of the $\pi$-$N$ scattering \cite{Pav02}, $f_{\pi} = 92 \, \mathrm{MeV}$ and the charged-pion mass $m_{\pi} = 140 \, \mathrm{MeV}$.
%
Then, Eq.~(\ref{eq:rel-123}) yields
\begin{equation}
\label{eq:cond-in-nucl}
\langle \overline{q} q \rangle_N \approx 0.5 \, \langle \overline{q} q \rangle,
\end{equation}
demonstrating a substantial suppression of the quark condensate in the nuclear matter. The value $\langle \overline{q} q \rangle_N$ can be interpreted as the sum of scalar densities of the $u$ (or $d$) quarks in vacuum and inside nucleons. The nucleon component of $\langle \overline{q} q \rangle_N$ is estimated in Ref.~\cite{Kovalenko:2013eba} to be $\approx (100 \, \mathrm{MeV})^3$. The sign of the nucleon component is opposite to the sign of the vacuum component; the latter is also numerically higher.

\section{The $0\nu\beta\beta$-decay rate and nuclear matrix elements}
\label{sec:NME}

The  $\beta$-decay Hamiltonian contains standard model
and non-standard neutrino interactions
\begin{eqnarray}
  {H}^{\beta} &=&~ \frac{G_{\beta}}{\sqrt{2}} ~
  \bar{e} \gamma^\rho (1-\gamma_5) \nu_{e}~\bar{u} \gamma_\rho (1-\gamma_5) d \\
  &&+ \frac{G_{\beta}}{\sqrt{2}} ~\frac{\varepsilon_{ee}}{4}~
  \bar{e} (1+\gamma_5) \nu^C_{e}~\bar{u} (1-\gamma_5) d ~+~h.c.\nonumber
\end{eqnarray}
Here, $G_\beta = G_F \cos{\theta_C}$, where $\cos{\theta_C}$ is the Cabbibo angle.

\begin{table*}[!t]
\centering
\caption{\label{tab:nme}
Nuclear matrix elements ${M}^\varepsilon$ (column 3), ${M}^\nu$ (column 4) and their ratio $f_{\rm nme}= g_A M^\nu/M^\epsilon$
(column 5) calculated within the Quasiparticle Random-Phase Approximation (QRPA)
method with partial restoration of the isospin symmetry and Argonne~V18 nucleon-nucleon potential \cite{Simkovic:2013qiy}.
${M}^\nu$ and $f_{\rm nme}$ are given for the unquenched value of the axial-vector coupling constant $g_A = 1.27$ 
In columns 6 and 7 the lower experimental limit on the $0\nu\beta\beta$-decay half-life and the upper constraint on the
$\varepsilon_{ee}$ are presented, respectively.  The phase-space factors $G^{0 \nu}$ (column 2) are  taken from
Ref.~\cite{Stefanik:2015twa}.
}
\renewcommand{\arraystretch}{1.2}
\begin{ruledtabular}
\begin{tabular}{ccccccc}
  Nucl. & $G^{0 \nu}$ [${\rm yr}^{-1}$] & ${M}^\varepsilon$ &  ${M}^\nu$ & $f_{\rm nme}$ & $T^{0\nu-{\rm exp}}_{1/2} [yr] $  Ref.  & $|\varepsilon_{ee}|$ \\\hline
  $\ce{^{76}Ge}$ & $0.237\times 10^{-14}$ & 5140  &  5.16  &  $1.27\times 10^{-3}$  & $> 8.0\times 10^{25}$  \cite{Geexp}  & $< 3.52 \times 10^{-10}$ \\
  $\ce{^{82}Se}$ & $1.018\times 10^{-14}$ & 4702  &  4.64  &  $1.25\times 10^{-3}$  & $> 2.4\times 10^{24}$  \cite{Seexp}  & $< 1.07 \times 10^{-9}$ \\
  $\ce{^{100}Mo}$ & $1.595\times 10^{-14}$ & 5751  &  5.40  & $1.19\times 10^{-3}$  & $> 1.1\times 10^{24}$  \cite{Moexp}  & $< 1.03 \times 10^{-9}$ \\
  $\ce{^{116}Cd}$ & $1.673\times 10^{-14}$ & 3232  &  4.04  & $1.59\times 10^{-3}$  & $> 2.2\times 10^{23}$  \cite{Cdexp}  & $< 4.02 \times 10^{-9}$ \\
  $\ce{^{130}Te}$ & $1.425\times 10^{-14}$ & 4530  &  3.89  & $1.09\times 10^{-3}$  & $> 3.2\times 10^{25}$  \cite{Teexp}  & $< 2.57 \times 10^{-10}$ \\
  $\ce{^{136}Xe}$ & $1.462\times 10^{-14}$ & 2530  &  2.18  & $1.09\times 10^{-3}$  & $> 1.07\times 10^{26}$ \cite{Gan16}  & $< 2.49 \times 10^{-10}$ \\
\end{tabular}
\end{ruledtabular}
\end{table*}

By the higher order perturbation of strong and electromagnetic interactions 
quark currents are converted into the ones involving nucleons. We  have
\begin{eqnarray}
  \label{nucleonme}
  && \langle n(p')| \overline{d} \gamma^\alpha (1-\gamma_5) u |p(p)\rangle =
\overline{n}(p')\left[ g_V(q^2) \gamma^\alpha \right. \\
  &&~~~~~~~ \left. + i g_M(q^2) \frac{\sigma^{\alpha\beta}}{2 m_p} q_\beta 
  -g_A(q^2) \gamma^\alpha\gamma_5 - g_P (q^2) q^\alpha \gamma_5\right] p(p), \nonumber\\
&& \langle n(p')| \overline{d} (1-\gamma_5) u |p(p)\rangle = 
\overline{n}(p')\left[ g_S(q^2) - g_{PS} (q^2) \gamma_5\right] p(p),\nonumber
\end{eqnarray} 
where $m_p$ is the nucleon mass, $q_\mu = (p'-p)_\mu$ is the momentum transfer and 
$p'$ and $p$ are the four momenta of neutron and proton, respectively. For the nucleon form factors
$g_V(q^2)$, $g_M(q^2)$, $g_A(q^2)$ and $g_P(q^2)$ we use parametrization as follows \cite{graf18}:
\begin{eqnarray}
  \frac{g_{V,M,S}(q^2)}{g_{V,M,S}} &=&  \left(1 + \frac{q^2}{m_V^2}\right)^{-2},~~
  \frac{g_{A}(q^2)}{g_{A}} = \left(1 + \frac{q^2}{m_A^2}\right)^{-2},\nonumber\\
  \frac{g_{PS}(q^2)}{g_{PS}} &=&   \left(1 + \frac{q^2}{m_A^2}\right)^{-2}
  \left(1+ \frac{q^2}{m_\pi^2}\right)^{-1}.\nonumber\\    
\end{eqnarray}
Here, $m_\pi$ is the mass of pion,
$q^2=\mathbf{q}\cdot\mathbf{q}$ (a small energy transfer in the nucleon vertex can be  neglected),
$m_V = 0.84$ GeV and $m_A = 1.09$ GeV and the renormalization constants: $g_V=1$, $g_A=1.269$,
$g_M=(\mu_p-\mu_n)$ = 3.70, $g_S = 1.0$ \cite{cirig18,Gonzalez14} and $g_{PS} = 349$ \cite{Gonzalez14}.  
The induced pseudoscalar coupling is given by the PCAC relation
\begin{eqnarray}
  g_{P}(q^2) = \frac{2 m_p}{{q^2}+ m^2_\pi} g_A(q^2). 
\end{eqnarray} 

To obtain nuclear matrix elements of interest, non-relativistic expansion of 
nucleon matrix elements in Eq. (\ref{nucleonme}) have to be performed.
For nuclear currents we get 
\begin{eqnarray}
  J_{V-A}^{\mu\dagger}(\mathbf{x})
  &=& \sum_{n=1}^A \tau^+_n \left[g^{\mu 0} J^0_{V-A}(\mathbf{q}) + g^{\mu k} J^k_{V-A}(\mathbf{q})\right]
  \delta(\mathbf{x}-\mathbf{r}_n), \nonumber\\
  J_{S-P}^{\dagger}(\mathbf{x})
  &=& \sum_{n=1}^A \tau^+_n J_{SP}(\mathbf{q}) \delta(\mathbf{x}-\mathbf{r}_n) 
\end{eqnarray} 
with $k=1,2,3$ and
\begin{eqnarray}
  J^0_{V-A}(\mathbf{q}) &=& g_V(q^2), \nonumber\\
  \mathbf{J}_{V-A}(\mathbf{q}) &=& g_M(q^2) i \frac{\bm{\sigma}\times\mathbf{q}}{2 m_p}
  - g_A(q^2) \bm{\sigma} + g_P(q^2) \frac{\mathbf{q}~\bm{\sigma}\cdot\mathbf{q}}{2 m_p}, \nonumber\\
  J_{S-P}(\mathbf{q}) &=& g_S(q^2) - g_{PS}(q^2)
  \frac{\bm{\sigma}\cdot\mathbf{q}}{2 m_p}, 
\end{eqnarray} 
where $\mathbf{r}_n$ is the coordinate of the $n$-th nucleon.

The nucleon recoil terms associated with the initial and final vertices in 
the $0\nu\beta\beta$-decay transition contain the nucleon recoil momenta
$\mathbf{q}_m$ and $\mathbf{q}_n$, respectively. They are opposite in direction
and their absolute values  $q_m=|\mathbf{q}_m|$ and $q_m=|\mathbf{q}_n|$ are roughly
equal to the absolute value of neutrino momentum $p=|\mathbf{p}|$. We have  \cite{Doi:1985dx}
\begin{eqnarray}
  \label{mutcanc}
 \mathbf{q}_m \simeq - \mathbf{q}_n \simeq \mathbf{p}.  
\end{eqnarray} 
In the amplitude only linear term in $\varepsilon_{ee}$ is considered. 
The main contribution to corresponding nuclear matrix element for ground state to ground state
$0^+\rightarrow 0^+$ transition is given by combinations of $g_{PS}$ term with $g_A$ and $g_P$
terms of nucleon currents and the spatial component of neutrino propagator proportional to neutrino
momentum $\mathbf{p}$.

For the inverse $0\nu\beta\beta$-decay half-life we obtain
\begin{eqnarray}
  \frac{1}{T_{1/2}} &=& G^{0\nu} 
 \left|\varepsilon_{ee} g_A M^\varepsilon + \frac{m_{\beta\beta}}{m_e} g_A^2 {M}^{\nu}\right|^2
\end{eqnarray}
Here, $G^{0\nu}$ is the known phase-space factor. 
Nuclear matrix elements ${M}^{\nu}$ and ${M}^{\varepsilon}$,
depend  on the nuclear structure of the particular isotopes 
$(A,Z)$, $(A,Z+1)$ and $(A,Z+2)$ under study.
The explicit form of ${M}^{\nu}$ can be found, e.g.,
in \cite{Simkovic:2007vu,Simkovic:2013qiy} and ${M}^{\varepsilon}$ is presented in a similar way here.

The nuclear matrix element $\mathcal{M}^{\varepsilon}$ consists of the Gamow--Teller (GT) and Tensor (T) parts
\begin{align}
\mathcal{M}^{\varepsilon} & = \mathcal{M}_{\mathrm{GT}}^{\varepsilon} + \mathcal{M}_{\mathrm{T}}^{\varepsilon}.
\end{align}

In the Quasiparticle Random-Phase Approximation (QRPA) method, $\mathcal{M}^{\nu, \varepsilon}$ is written via sums over the virtual intermediate states labeled by their angular momenta and parities $J^{\pi}$ and indices $k_i$ and $k_f$ \cite{Simkovic:2007vu,Simkovic:2013qiy}
\begin{align}
\label{eq:long}
& \mathcal{M}_K = \sum_{J^{\pi}, k_i, k_f, \mathcal{J}} \sum_{p n p' n'} (-1)^{j_n + j_{p'} + J + \mathcal{J}} \sqrt{2 \mathcal{J} + 1} \nonumber \\
& \times
\begin{Bmatrix}
j_p & j_n & J \\
j_{n'} & j_{p'} & \mathcal{J}
\end{Bmatrix}
\braket{p(1), \, p'(2); \, \mathcal{J} \| \mathcal{O}_K \| n(1), \, n'(2); \, \mathcal{J}} \nonumber \\
& \times \braket{0_f^+ \| [\widetilde{c_{p'}^{\dagger} \tilde{c}_{n'}}]_J \| J^{\pi} k_f} \braket{J^{\pi} k_f | J^{\pi} k_i} \braket{J^{\pi} k_fi \| [c_p^{\dagger} \tilde{c}_n]_J \| 0_i^+}.
\end{align}
The reduced matrix elements of the one-body operators $c_p^{\dagger} \tilde{c}_n$ ($\tilde{c}_n$ denotes
the time-reversed state) in the Eq.~(\ref{eq:long}) depend on the BCS coefficients $u_i$, $v_j$ and
on the QRPA vectors $X$, $Y$ \cite{Simkovic:2013qiy}.

The two-body operators $O^{\varepsilon}_K$ in (\ref{eq:long}), where $K = \mathrm{GT}, \, \mathrm{T} \textnormal{ (Tensor)}$,
contain neutrino potentials, spin and isospin operators, and RPA energies $E_{J^{\pi}}^{k_i, k_f}$:
\begin{align}
\label{a12}
O^{\varepsilon}_{\mathrm{GT}}(r_{12}, \, E_{J^{\pi}}^k) & = \tau^+(1) \, \tau^+(2) \, H^{\varepsilon}_{\mathrm{GT}}(r_{12}, \, E_{J^{\pi}}^k) \, \sigma_{12}, \nonumber \\
O^{\varepsilon}_{\mathrm{T}}(r_{12}, \, E_{J^{\pi}}^k) & = \tau^+(1) \, \tau^+(2) \, H^{\varepsilon}_{\mathrm{T}}(r_{12}, \, E_{J^{\pi}}^k) \, S_{12}
\end{align}
with
\begin{align}
\mathbf{r_{12}} & = \mathbf{r_1} - \mathbf{r_2}, \quad r_{12} \equiv |\mathbf{r_{12}}|, \quad \mathbf{\hat{r}_{12}} \equiv \frac{\mathbf{r_{12}}}{r_{12}}, \nonumber \\
\sigma_{12} & = \bm{\sigma_1} \cdot \bm{\sigma_2}, \nonumber \\
S_{12} & = 3 \, (\bm{\sigma_1} \cdot \mathbf{\hat{r}_{12}}) (\bm{\sigma_2} \cdot \mathbf{\hat{r}_{12}}) - \sigma_{12}.
\end{align}
Here, $\mathbf{r_1}$ and $\mathbf{r_2}$ are the coordinates nucleons undergoing the beta decay.

The neutrino potentials are
\begin{align}
\label{eq:pot}
& H_K^{\varepsilon}(r_{12}, \, E_{J^{\pi}}^k) = \frac{2}{\pi} \, R \nonumber \\
& \times \int\limits_0^{\infty} f_K(p r_{12}) \, \frac{h^{\varepsilon}(p^2) \, p \, \mathrm{d}p}{p + E_{J^{\pi}}^k - \frac{1}{2} (E_i + E_f)},
\end{align}
where $f_{\mathrm{F}, \mathrm{GT}}(q r_{12}) = j_0(q r_{12})$ and $f_{\mathrm{T}}(q r_{12}) = -j_2(q r_{12})$ are the spherical Bessel functions. The potentials (\ref{eq:pot}) depend explicitly---though rather 
weakly---on the energies $E_{J^{\pi}}^k$ of the virtual intermediate states. The functions $h(p^2)$
in Eq.~(\ref{eq:pot}) is defined as
\begin{align}
h^{\varepsilon}(p^2) & = \frac{1}{12} \, \frac{F_P^{(3)}(p^2) \, g_A(p^2)}{g_A} \left( 1 - \frac{p^2}{p^2 + m_{\pi}^2} \right) \frac{p^2}{m_p m_{e}}.
\end{align}

In Table~\ref{tab:nme}, we show nuclear matrix elements
${M}^\varepsilon$, ${M}^\nu$ and their ratio $f_{\rm NME}$ 
calculated within the QRPA method with partial restoration of the isospin symmetry.
Details of the calculation are given in \cite{Simkovic:2013qiy}.
By glancing Table~\ref{tab:nme} we see that ${M}^\varepsilon$
is by about factor 200 larger when compared with  ${M}^\nu$, what is
a result of the  additional factor $p/(2 m_e)$ in the neutrino exchange potential
(see Eq. (\ref{eq:pot})). It is worth noting that within considered approximations
${M}^\varepsilon$ does not depend on axial-vector coupling constant $g_A$.
The most stringent current experimental limits on the $0\nu\beta\beta$-decay half-life 
and corresponding upper bounds on the absolute value of $\varepsilon_{ee}$ are presented
in Table \ref{tab:nme} as well.

\bibliographystyle{apsrev}
%
\bibliography{bibliography}

\begin{thebibliography}{45}
\expandafter\ifx\csname natexlab\endcsname\relax\def\natexlab#1{#1}\fi
\expandafter\ifx\csname bibnamefont\endcsname\relax
  \def\bibnamefont#1{#1}\fi
\expandafter\ifx\csname bibfnamefont\endcsname\relax
  \def\bibfnamefont#1{#1}\fi
\expandafter\ifx\csname citenamefont\endcsname\relax
  \def\citenamefont#1{#1}\fi
\expandafter\ifx\csname url\endcsname\relax
  \def\url#1{\texttt{#1}}\fi
\expandafter\ifx\csname urlprefix\endcsname\relax\def\urlprefix{URL }\fi
\providecommand{\bibinfo}[2]{#2}
\providecommand{\eprint}[2][]{\url{#2}}

\bibitem[{\citenamefont{Cai et~al.}(2017)\citenamefont{Cai,
  Herrero-Garc{\'\i}a, Schmidt, Vicente, and Volkas}}]{Cai:2017jrq}
\bibinfo{author}{\bibfnamefont{Y.}~\bibnamefont{Cai}},
  \bibinfo{author}{\bibfnamefont{J.}~\bibnamefont{Herrero-Garc{\'\i}a}},
  \bibinfo{author}{\bibfnamefont{M.~A.} \bibnamefont{Schmidt}},
  \bibinfo{author}{\bibfnamefont{A.}~\bibnamefont{Vicente}}, \bibnamefont{and}
  \bibinfo{author}{\bibfnamefont{R.~R.} \bibnamefont{Volkas}},
  \bibinfo{journal}{Front.in Phys.} \textbf{\bibinfo{volume}{5}},
  \bibinfo{pages}{63} (\bibinfo{year}{2017}), \eprint{1706.08524}.

\bibitem[{\citenamefont{Ma and Popov}(2017)}]{Ma:2016mwh}
\bibinfo{author}{\bibfnamefont{E.}~\bibnamefont{Ma}} \bibnamefont{and}
  \bibinfo{author}{\bibfnamefont{O.}~\bibnamefont{Popov}},
  \bibinfo{journal}{Phys. Lett.} \textbf{\bibinfo{volume}{B764}},
  \bibinfo{pages}{142} (\bibinfo{year}{2017}), \eprint{1609.02538}.

\bibitem[{\citenamefont{Yao and Ding}(2017)}]{Yao:2017vtm}
\bibinfo{author}{\bibfnamefont{C.-Y.} \bibnamefont{Yao}} \bibnamefont{and}
  \bibinfo{author}{\bibfnamefont{G.-J.} \bibnamefont{Ding}},
  \bibinfo{journal}{Phys. Rev.} \textbf{\bibinfo{volume}{D96}},
  \bibinfo{pages}{095004} (\bibinfo{year}{2017}), \bibinfo{note}{[Erratum:
  Phys. Rev.D98,no.3,039901(2018)]}, \eprint{1707.09786}.

\bibitem[{\citenamefont{Centelles~Chuli\'a
  et~al.}(2018)\citenamefont{Centelles~Chuli\'a, Srivastava, and
  Valle}}]{CentellesChulia:2018bkz}
\bibinfo{author}{\bibfnamefont{S.}~\bibnamefont{Centelles~Chuli\'a}},
  \bibinfo{author}{\bibfnamefont{R.}~\bibnamefont{Srivastava}},
  \bibnamefont{and} \bibinfo{author}{\bibfnamefont{J.~W.~F.}
  \bibnamefont{Valle}}, \bibinfo{journal}{Phys. Rev.}
  \textbf{\bibinfo{volume}{D98}}, \bibinfo{pages}{035009}
  (\bibinfo{year}{2018}), \eprint{1804.03181}.

\bibitem[{\citenamefont{Centelles~Chuli\'a
  et~al.}(2019)\citenamefont{Centelles~Chuli\'a, Cepedello, Peinado, and
  Srivastava}}]{CentellesChulia:2019xky}
\bibinfo{author}{\bibfnamefont{S.}~\bibnamefont{Centelles~Chuli\'a}},
  \bibinfo{author}{\bibfnamefont{R.}~\bibnamefont{Cepedello}},
  \bibinfo{author}{\bibfnamefont{E.}~\bibnamefont{Peinado}}, \bibnamefont{and}
  \bibinfo{author}{\bibfnamefont{R.}~\bibnamefont{Srivastava}}
  (\bibinfo{year}{2019}), \eprint{1907.08630}.

\bibitem[{\citenamefont{Arbel\'{a}ez et~al.}(2019)\citenamefont{Arbel\'{a}ez,
  C\'arcamo, Cepedello, Hirsch, and Kovalenko}}]{Arbelaez:2019wyz}
\bibinfo{author}{\bibfnamefont{C.}~\bibnamefont{Arbel\'{a}ez}},
  \bibinfo{author}{\bibfnamefont{A.~E.} \bibnamefont{C\'arcamo}},
  \bibinfo{author}{\bibfnamefont{R.}~\bibnamefont{Cepedello}},
  \bibinfo{author}{\bibfnamefont{M.}~\bibnamefont{Hirsch}}, \bibnamefont{and}
  \bibinfo{author}{\bibfnamefont{S.}~\bibnamefont{Kovalenko}}
  (\bibinfo{year}{2019}), \eprint{1910.04178}.

\bibitem[{\citenamefont{Arbel\'aez et~al.}(2019)\citenamefont{Arbel\'aez,
  C\'arcamo~Hern\'andez, Cepedello, Kovalenko, and Schmidt}}]{Arbelaez:2019ofg}
\bibinfo{author}{\bibfnamefont{C.}~\bibnamefont{Arbel\'aez}},
  \bibinfo{author}{\bibfnamefont{A.~E.} \bibnamefont{C\'arcamo~Hern\'andez}},
  \bibinfo{author}{\bibfnamefont{R.}~\bibnamefont{Cepedello}},
  \bibinfo{author}{\bibfnamefont{S.}~\bibnamefont{Kovalenko}},
  \bibnamefont{and} \bibinfo{author}{\bibfnamefont{I.}~\bibnamefont{Schmidt}}
  (\bibinfo{year}{2019}), \eprint{1911.02033}.

\bibitem[{\citenamefont{Babu et~al.}(2020)\citenamefont{Babu, Bhupal~Dev, Jana,
  and Thapa}}]{Babu20}
\bibinfo{author}{\bibfnamefont{K.~S.} \bibnamefont{Babu}},
  \bibinfo{author}{\bibfnamefont{P.~S.} \bibnamefont{Bhupal~Dev}},
  \bibinfo{author}{\bibfnamefont{S.}~\bibnamefont{Jana}}, \bibnamefont{and}
  \bibinfo{author}{\bibfnamefont{A.}~\bibnamefont{Thapa}}, \bibinfo{journal}{J.
  High Energy Phys.} \textbf{\bibinfo{volume}{03}}, \bibinfo{pages}{006}
  (\bibinfo{year}{2020}).

\bibitem[{\citenamefont{P\"{a}s et~al.}(1999)\citenamefont{P\"{a}s, Hirsch,
  Klapdor-Kleingrothaus, and Kovalenko}}]{Pas:1999fc}
\bibinfo{author}{\bibfnamefont{H.}~\bibnamefont{P\"{a}s}},
  \bibinfo{author}{\bibfnamefont{M.}~\bibnamefont{Hirsch}},
  \bibinfo{author}{\bibfnamefont{H.~V.} \bibnamefont{Klapdor-Kleingrothaus}},
  \bibnamefont{and} \bibinfo{author}{\bibfnamefont{S.~G.}
  \bibnamefont{Kovalenko}}, \bibinfo{journal}{Phys. Lett. B}
  \textbf{\bibinfo{volume}{453}}, \bibinfo{pages}{194} (\bibinfo{year}{1999}).

\bibitem[{\citenamefont{Deppisch et~al.}(2012)\citenamefont{Deppisch, Hirsch,
  and P\"{a}s}}]{Deppisch:2012nb}
\bibinfo{author}{\bibfnamefont{F.~F.} \bibnamefont{Deppisch}},
  \bibinfo{author}{\bibfnamefont{M.}~\bibnamefont{Hirsch}}, \bibnamefont{and}
  \bibinfo{author}{\bibfnamefont{H.}~\bibnamefont{P\"{a}s}},
  \bibinfo{journal}{J.Phys.} \textbf{\bibinfo{volume}{G39}},
  \bibinfo{pages}{124007} (\bibinfo{year}{2012}), \eprint{1208.0727}.

\bibitem[{\citenamefont{Arbel\'{a}ez et~al.}(2016)\citenamefont{Arbel\'{a}ez,
  Gonz\'{a}lez, Hirsch, and Kovalenko}}]{Arbelaez:2016zlt}
\bibinfo{author}{\bibfnamefont{C.}~\bibnamefont{Arbel\'{a}ez}},
  \bibinfo{author}{\bibfnamefont{M.}~\bibnamefont{Gonz\'{a}lez}},
  \bibinfo{author}{\bibfnamefont{M.}~\bibnamefont{Hirsch}}, \bibnamefont{and}
  \bibinfo{author}{\bibfnamefont{S.}~\bibnamefont{Kovalenko}},
  \bibinfo{journal}{Phys. Rev.} \textbf{\bibinfo{volume}{D94}},
  \bibinfo{pages}{096014} (\bibinfo{year}{2016}), \bibinfo{note}{[erratum:
  Phys. Rev.D97,no.9,099904(2018)]}, \eprint{1610.04096}.

\bibitem[{\citenamefont{Cirigliano et~al.}(2017)\citenamefont{Cirigliano,
  Dekens, de~Vries, Graesser, and Mereghetti}}]{Cirigliano:2017djv}
\bibinfo{author}{\bibfnamefont{V.}~\bibnamefont{Cirigliano}},
  \bibinfo{author}{\bibfnamefont{W.}~\bibnamefont{Dekens}},
  \bibinfo{author}{\bibfnamefont{J.}~\bibnamefont{de~Vries}},
  \bibinfo{author}{\bibfnamefont{M.}~\bibnamefont{Graesser}}, \bibnamefont{and}
  \bibinfo{author}{\bibfnamefont{E.}~\bibnamefont{Mereghetti}},
  \bibinfo{journal}{JHEP} \textbf{\bibinfo{volume}{12}}, \bibinfo{pages}{082}
  (\bibinfo{year}{2017}), \eprint{1708.09390}.

\bibitem[{\citenamefont{Thomas and Xu}(1992)}]{Tho92}
\bibinfo{author}{\bibfnamefont{S.~D.} \bibnamefont{Thomas}} \bibnamefont{and}
  \bibinfo{author}{\bibfnamefont{R.-M.} \bibnamefont{Xu}},
  \bibinfo{journal}{Phys. Lett. B} \textbf{\bibinfo{volume}{284}},
  \bibinfo{pages}{341} (\bibinfo{year}{1992}).

\bibitem[{\citenamefont{McNeile et~al.}(2013)\citenamefont{McNeile, Bazavov,
  Davies, Dowdall, Hornbostel, Lepage, and Trottier}}]{McNeile:2012xh}
\bibinfo{author}{\bibfnamefont{C.}~\bibnamefont{McNeile}},
  \bibinfo{author}{\bibfnamefont{A.}~\bibnamefont{Bazavov}},
  \bibinfo{author}{\bibfnamefont{C.~T.~H.} \bibnamefont{Davies}},
  \bibinfo{author}{\bibfnamefont{R.~J.} \bibnamefont{Dowdall}},
  \bibinfo{author}{\bibfnamefont{K.}~\bibnamefont{Hornbostel}},
  \bibinfo{author}{\bibfnamefont{G.~P.} \bibnamefont{Lepage}},
  \bibnamefont{and} \bibinfo{author}{\bibfnamefont{H.~D.}
  \bibnamefont{Trottier}}, \bibinfo{journal}{Phys. Rev.}
  \textbf{\bibinfo{volume}{D87}}, \bibinfo{pages}{034503}
  (\bibinfo{year}{2013}), \eprint{1211.6577}.

\bibitem[{\citenamefont{Allton et~al.}(2008)}]{Allton:2008pn}
\bibinfo{author}{\bibfnamefont{C.}~\bibnamefont{Allton}} \bibnamefont{et~al.}
  (\bibinfo{collaboration}{RBC-UKQCD}), \bibinfo{journal}{Phys. Rev.}
  \textbf{\bibinfo{volume}{D78}}, \bibinfo{pages}{114509}
  (\bibinfo{year}{2008}), \eprint{0804.0473}.

\bibitem[{\citenamefont{Kitazawa and Sakai}(2018)}]{Kitazawa:2017hqk}
\bibinfo{author}{\bibfnamefont{N.}~\bibnamefont{Kitazawa}} \bibnamefont{and}
  \bibinfo{author}{\bibfnamefont{Y.}~\bibnamefont{Sakai}},
  \bibinfo{journal}{Int. J. Mod. Phys.} \textbf{\bibinfo{volume}{A33}},
  \bibinfo{pages}{1850017} (\bibinfo{year}{2018}), \eprint{1707.03559}.

\bibitem[{\citenamefont{Davoudiasl and Everett}(2006)}]{Davoudiasl:2005ai}
\bibinfo{author}{\bibfnamefont{H.}~\bibnamefont{Davoudiasl}} \bibnamefont{and}
  \bibinfo{author}{\bibfnamefont{L.~L.} \bibnamefont{Everett}},
  \bibinfo{journal}{Phys. Lett.} \textbf{\bibinfo{volume}{B634}},
  \bibinfo{pages}{55} (\bibinfo{year}{2006}), \eprint{hep-ph/0512188}.

\bibitem[{Note1()}]{Note1}
Note1, \bibinfo{note}{we are thankful to Martin Hirsch for drawing our
  attention to this fact.}

\bibitem[{\citenamefont{de~Salas et~al.}(2018)\citenamefont{de~Salas, Forero,
  Ternes, T\'{o}rtola, and Valle}}]{deSalas:2017kay}
\bibinfo{author}{\bibfnamefont{P.~F.} \bibnamefont{de~Salas}},
  \bibinfo{author}{\bibfnamefont{D.~V.} \bibnamefont{Forero}},
  \bibinfo{author}{\bibfnamefont{C.~A.} \bibnamefont{Ternes}},
  \bibinfo{author}{\bibfnamefont{M.}~\bibnamefont{T\'{o}rtola}},
  \bibnamefont{and} \bibinfo{author}{\bibfnamefont{J.~W.~F.}
  \bibnamefont{Valle}}, \bibinfo{journal}{Phys. Lett.}
  \textbf{\bibinfo{volume}{B782}}, \bibinfo{pages}{633} (\bibinfo{year}{2018}),
  \eprint{1708.01186}.

\bibitem[{\citenamefont{Aghanim et~al.}(2018)}]{Agh18}
\bibinfo{author}{\bibfnamefont{N.}~\bibnamefont{Aghanim}} \bibnamefont{et~al.}
  (\bibinfo{collaboration}{Planck}) (\bibinfo{year}{2018}),
  \eprint{1807.06209}.

\bibitem[{\citenamefont{Gando et~al.}(2016)}]{Gan16}
\bibinfo{author}{\bibfnamefont{A.}~\bibnamefont{Gando}} \bibnamefont{et~al.}
  (\bibinfo{collaboration}{KamLAND-Zen}), \bibinfo{journal}{Phys. Rev. Lett.}
  \textbf{\bibinfo{volume}{117}}, \bibinfo{pages}{082503}
  (\bibinfo{year}{2016}).

\bibitem[{\citenamefont{Vagnozzi et~al.}(2017)}]{Vagno17}
\bibinfo{author}{\bibfnamefont{S.}~\bibnamefont{Vagnozzi}}
  \bibnamefont{et~al.}, \bibinfo{journal}{Phys. Rev. D}
  \textbf{\bibinfo{volume}{96}}, \bibinfo{pages}{123503}
  (\bibinfo{year}{2017}).

\bibitem[{\citenamefont{Kovalenko et~al.}(2014)\citenamefont{Kovalenko,
  Krivoruchenko, and \v{S}imkovic}}]{Kovalenko:2013eba}
\bibinfo{author}{\bibfnamefont{S.}~\bibnamefont{Kovalenko}},
  \bibinfo{author}{\bibfnamefont{M.~I.} \bibnamefont{Krivoruchenko}},
  \bibnamefont{and}
  \bibinfo{author}{\bibfnamefont{F.}~\bibnamefont{\v{S}imkovic}},
  \bibinfo{journal}{Phys. Rev. Lett.} \textbf{\bibinfo{volume}{112}},
  \bibinfo{pages}{142503} (\bibinfo{year}{2014}), \eprint{1311.4200}.

\bibitem[{\citenamefont{Helo et~al.}(2011)\citenamefont{Helo, Kovalenko, and
  Schmidt}}]{Helo:2010cw}
\bibinfo{author}{\bibfnamefont{J.~C.} \bibnamefont{Helo}},
  \bibinfo{author}{\bibfnamefont{S.}~\bibnamefont{Kovalenko}},
  \bibnamefont{and} \bibinfo{author}{\bibfnamefont{I.}~\bibnamefont{Schmidt}},
  \bibinfo{journal}{Nucl. Phys. B} \textbf{\bibinfo{volume}{853}},
  \bibinfo{pages}{80} (\bibinfo{year}{2011}), \eprint{1005.1607}.

\bibitem[{\citenamefont{Drexlin et~al.}(2013)\citenamefont{Drexlin, Hannen,
  Mertens, and Weinheimer}}]{Drexlin:2013lha}
\bibinfo{author}{\bibfnamefont{G.}~\bibnamefont{Drexlin}},
  \bibinfo{author}{\bibfnamefont{V.}~\bibnamefont{Hannen}},
  \bibinfo{author}{\bibfnamefont{S.}~\bibnamefont{Mertens}}, \bibnamefont{and}
  \bibinfo{author}{\bibfnamefont{C.}~\bibnamefont{Weinheimer}},
  \bibinfo{journal}{Adv. High Energy Phys.} \textbf{\bibinfo{volume}{2013}},
  \bibinfo{pages}{293986} (\bibinfo{year}{2013}), \eprint{1307.0101}.

\bibitem[{\citenamefont{Aker et~al.}(2019)}]{Aker:2019uuj}
\bibinfo{author}{\bibfnamefont{M.}~\bibnamefont{Aker}} \bibnamefont{et~al.}
  (\bibinfo{collaboration}{KATRIN}) (\bibinfo{year}{2019}),
  \eprint{1909.06048}.

\bibitem[{\citenamefont{C\'{a}rcamo~Hern\'{a}ndez
  et~al.}(2017)\citenamefont{C\'{a}rcamo~Hern\'{a}ndez, Kovalenko, and
  Schmidt}}]{CarcamoHernandez:2016pdu}
\bibinfo{author}{\bibfnamefont{A.~E.} \bibnamefont{C\'{a}rcamo~Hern\'{a}ndez}},
  \bibinfo{author}{\bibfnamefont{S.}~\bibnamefont{Kovalenko}},
  \bibnamefont{and} \bibinfo{author}{\bibfnamefont{I.}~\bibnamefont{Schmidt}},
  \bibinfo{journal}{JHEP} \textbf{\bibinfo{volume}{02}}, \bibinfo{pages}{125}
  (\bibinfo{year}{2017}), \eprint{1611.09797}.

\bibitem[{\citenamefont{Capozzi et~al.}(2016)\citenamefont{Capozzi, Lisi,
  Marrone, Montanino, and Palazzo}}]{Cap16}
\bibinfo{author}{\bibfnamefont{F.}~\bibnamefont{Capozzi}},
  \bibinfo{author}{\bibfnamefont{E.}~\bibnamefont{Lisi}},
  \bibinfo{author}{\bibfnamefont{A.}~\bibnamefont{Marrone}},
  \bibinfo{author}{\bibfnamefont{D.}~\bibnamefont{Montanino}},
  \bibnamefont{and} \bibinfo{author}{\bibfnamefont{A.}~\bibnamefont{Palazzo}},
  \bibinfo{journal}{Nucl. Phys. B} \textbf{\bibinfo{volume}{908}},
  \bibinfo{pages}{218} (\bibinfo{year}{2016}).

\bibitem[{\citenamefont{Cohen et~al.}(1992)\citenamefont{Cohen, Furnstahl, and
  Griegel}}]{Coh92}
\bibinfo{author}{\bibfnamefont{T.~D.} \bibnamefont{Cohen}},
  \bibinfo{author}{\bibfnamefont{R.~J.} \bibnamefont{Furnstahl}},
  \bibnamefont{and} \bibinfo{author}{\bibfnamefont{D.~K.}
  \bibnamefont{Griegel}}, \bibinfo{journal}{Phys. Rev. C}
  \textbf{\bibinfo{volume}{45}}, \bibinfo{pages}{1881} (\bibinfo{year}{1992}).

\bibitem[{\citenamefont{Tarrach}(1982)}]{Tarrach:1981bi}
\bibinfo{author}{\bibfnamefont{R.}~\bibnamefont{Tarrach}},
  \bibinfo{journal}{Nucl. Phys.} \textbf{\bibinfo{volume}{B196}},
  \bibinfo{pages}{45} (\bibinfo{year}{1982}).

\bibitem[{\citenamefont{Jaffe and Korpa}(1987)}]{Jaffe:1987sw}
\bibinfo{author}{\bibfnamefont{R.~L.} \bibnamefont{Jaffe}} \bibnamefont{and}
  \bibinfo{author}{\bibfnamefont{C.~L.} \bibnamefont{Korpa}},
  \bibinfo{journal}{Comments Nucl. Part. Phys.} \textbf{\bibinfo{volume}{17}},
  \bibinfo{pages}{163} (\bibinfo{year}{1987}).

\bibitem[{\citenamefont{Gell-Mann et~al.}(1968)\citenamefont{Gell-Mann, Oakes,
  and Renner}}]{Gel68}
\bibinfo{author}{\bibfnamefont{M.}~\bibnamefont{Gell-Mann}},
  \bibinfo{author}{\bibfnamefont{R.~J.} \bibnamefont{Oakes}}, \bibnamefont{and}
  \bibinfo{author}{\bibfnamefont{B.}~\bibnamefont{Renner}},
  \bibinfo{journal}{Phys. Rev.} \textbf{\bibinfo{volume}{175}},
  \bibinfo{pages}{2195} (\bibinfo{year}{1968}).

\bibitem[{\citenamefont{Pavan et~al.}(2002)\citenamefont{Pavan, Strakovsky,
  Workman, and Arndt}}]{Pav02}
\bibinfo{author}{\bibfnamefont{M.~M.} \bibnamefont{Pavan}},
  \bibinfo{author}{\bibfnamefont{I.~I.} \bibnamefont{Strakovsky}},
  \bibinfo{author}{\bibfnamefont{R.~L.} \bibnamefont{Workman}},
  \bibnamefont{and} \bibinfo{author}{\bibfnamefont{R.~A.} \bibnamefont{Arndt}},
  \bibinfo{journal}{PiN Newslett.} \textbf{\bibinfo{volume}{16}},
  \bibinfo{pages}{110} (\bibinfo{year}{2002}), \eprint{0111066}.

\bibitem[{\citenamefont{{\v S}imkovic et~al.}(2013)\citenamefont{{\v S}imkovic,
  Rodin, Faessler, and Vogel}}]{Simkovic:2013qiy}
\bibinfo{author}{\bibfnamefont{F.}~\bibnamefont{{\v S}imkovic}},
  \bibinfo{author}{\bibfnamefont{V.}~\bibnamefont{Rodin}},
  \bibinfo{author}{\bibfnamefont{A.}~\bibnamefont{Faessler}}, \bibnamefont{and}
  \bibinfo{author}{\bibfnamefont{P.}~\bibnamefont{Vogel}},
  \bibinfo{journal}{Phys.Rev.} \textbf{\bibinfo{volume}{C87}},
  \bibinfo{pages}{045501} (\bibinfo{year}{2013}), \eprint{1302.1509}.

\bibitem[{\citenamefont{\v{S}tef\'{a}nik
  et~al.}(2015)\citenamefont{\v{S}tef\'{a}nik, Dvornick\'{y}, \v{S}imkovic, and
  Vogel}}]{Stefanik:2015twa}
\bibinfo{author}{\bibfnamefont{D.}~\bibnamefont{\v{S}tef\'{a}nik}},
  \bibinfo{author}{\bibfnamefont{R.}~\bibnamefont{Dvornick\'{y}}},
  \bibinfo{author}{\bibfnamefont{F.}~\bibnamefont{\v{S}imkovic}},
  \bibnamefont{and} \bibinfo{author}{\bibfnamefont{P.}~\bibnamefont{Vogel}},
  \bibinfo{journal}{Phys. Rev.} \textbf{\bibinfo{volume}{C92}},
  \bibinfo{pages}{055502} (\bibinfo{year}{2015}), \eprint{1506.07145}.

\bibitem[{\citenamefont{Agostini et~al.}(2018)}]{Geexp}
\bibinfo{author}{\bibfnamefont{M.}~\bibnamefont{Agostini}} \bibnamefont{et~al.}
  (\bibinfo{collaboration}{GERDA}), \bibinfo{journal}{Phys. Rev. Lett.}
  \textbf{\bibinfo{volume}{120}}, \bibinfo{pages}{132503}
  (\bibinfo{year}{2018}).

\bibitem[{\citenamefont{Azzolini et~al.}(2018)}]{Seexp}
\bibinfo{author}{\bibfnamefont{O.}~\bibnamefont{Azzolini}} \bibnamefont{et~al.}
  (\bibinfo{collaboration}{CUPID-0}), \bibinfo{journal}{Phys. Rev. Lett.}
  \textbf{\bibinfo{volume}{120}}, \bibinfo{pages}{232502}
  (\bibinfo{year}{2018}).

\bibitem[{\citenamefont{Arnold et~al.}(2015)}]{Moexp}
\bibinfo{author}{\bibfnamefont{R.}~\bibnamefont{Arnold}} \bibnamefont{et~al.}
  (\bibinfo{collaboration}{NEMO-3}), \bibinfo{journal}{Phys. Rev. D}
  \textbf{\bibinfo{volume}{92}}, \bibinfo{pages}{072011}
  (\bibinfo{year}{2015}).

\bibitem[{\citenamefont{Tretyak et~al.}(2019)}]{Cdexp}
\bibinfo{author}{\bibfnamefont{V.~I.} \bibnamefont{Tretyak}}
  \bibnamefont{et~al.} (\bibinfo{collaboration}{Aurora}), \bibinfo{journal}{AIP
  Conf. Proc.} \textbf{\bibinfo{volume}{81}}, \bibinfo{pages}{020029}
  (\bibinfo{year}{2019}).

\bibitem[{\citenamefont{Adams et~al.}(2020)}]{Teexp}
\bibinfo{author}{\bibfnamefont{D.~Q.} \bibnamefont{Adams}} \bibnamefont{et~al.}
  (\bibinfo{collaboration}{CUORE}), \bibinfo{journal}{Phys. Rev. Lett.}
  \textbf{\bibinfo{volume}{124}}, \bibinfo{pages}{122501}
  (\bibinfo{year}{2020}).

\bibitem[{\citenamefont{Graf et~al.}(2018)\citenamefont{Graf, Deppisch,
  Iachello, and Kotila}}]{graf18}
\bibinfo{author}{\bibfnamefont{L.}~\bibnamefont{Graf}},
  \bibinfo{author}{\bibfnamefont{F.~F.} \bibnamefont{Deppisch}},
  \bibinfo{author}{\bibfnamefont{F.}~\bibnamefont{Iachello}}, \bibnamefont{and}
  \bibinfo{author}{\bibfnamefont{J.}~\bibnamefont{Kotila}},
  \bibinfo{journal}{Phys. Rev. D} \textbf{\bibinfo{volume}{98}},
  \bibinfo{pages}{095023} (\bibinfo{year}{2018}).

\bibitem[{\citenamefont{Gupta et~al.}(2018)\citenamefont{Gupta, Jang, Yoon,
  Lin, Cirigliano, and Bhattacharya}}]{cirig18}
\bibinfo{author}{\bibfnamefont{R.}~\bibnamefont{Gupta}},
  \bibinfo{author}{\bibfnamefont{Y.}~\bibnamefont{Jang}},
  \bibinfo{author}{\bibfnamefont{B.}~\bibnamefont{Yoon}},
  \bibinfo{author}{\bibfnamefont{H.}~\bibnamefont{Lin}},
  \bibinfo{author}{\bibfnamefont{V.}~\bibnamefont{Cirigliano}},
  \bibnamefont{and}
  \bibinfo{author}{\bibfnamefont{T.}~\bibnamefont{Bhattacharya}},
  \bibinfo{journal}{Phys. Rev. D} \textbf{\bibinfo{volume}{98}},
  \bibinfo{pages}{034503} (\bibinfo{year}{2018}).

\bibitem[{\citenamefont{Gonz\'{a}lez-Alonso and
  Martin~Camalich}(2014)}]{Gonzalez14}
\bibinfo{author}{\bibfnamefont{M.}~\bibnamefont{Gonz\'{a}lez-Alonso}}
  \bibnamefont{and}
  \bibinfo{author}{\bibfnamefont{J.}~\bibnamefont{Martin~Camalich}},
  \bibinfo{journal}{Phys. Rev. Lett.} \textbf{\bibinfo{volume}{112}},
  \bibinfo{pages}{042501} (\bibinfo{year}{2014}).

\bibitem[{\citenamefont{Doi et~al.}(1985)\citenamefont{Doi, Kotani, and
  Takasugi}}]{Doi:1985dx}
\bibinfo{author}{\bibfnamefont{M.}~\bibnamefont{Doi}},
  \bibinfo{author}{\bibfnamefont{T.}~\bibnamefont{Kotani}}, \bibnamefont{and}
  \bibinfo{author}{\bibfnamefont{E.}~\bibnamefont{Takasugi}},
  \bibinfo{journal}{Prog.Theor.Phys.Suppl.} \textbf{\bibinfo{volume}{83}},
  \bibinfo{pages}{1} (\bibinfo{year}{1985}).

\bibitem[{\citenamefont{\v{S}imkovic et~al.}(2008)\citenamefont{\v{S}imkovic,
  Faessler, Rodin, Vogel, and Engel}}]{Simkovic:2007vu}
\bibinfo{author}{\bibfnamefont{F.}~\bibnamefont{\v{S}imkovic}},
  \bibinfo{author}{\bibfnamefont{A.}~\bibnamefont{Faessler}},
  \bibinfo{author}{\bibfnamefont{V.}~\bibnamefont{Rodin}},
  \bibinfo{author}{\bibfnamefont{P.}~\bibnamefont{Vogel}}, \bibnamefont{and}
  \bibinfo{author}{\bibfnamefont{J.}~\bibnamefont{Engel}},
  \bibinfo{journal}{Phys. Rev.} \textbf{\bibinfo{volume}{C77}},
  \bibinfo{pages}{045503} (\bibinfo{year}{2008}), \eprint{0710.2055}.

\end{thebibliography}
%
\end{document}